\documentclass[conference]{IEEEtran}
\usepackage{array}
\usepackage{booktabs} 
\usepackage{amssymb,latexsym,amsmath}    
 \usepackage{booktabs}
\usepackage{multirow}
\usepackage{booktabs, multicol, multirow}
\usepackage{graphicx}
\usepackage{epstopdf}
\usepackage{enumerate}
\usepackage{tikz}
\usepackage{threeparttable}
\usepackage{siunitx}
\usepackage{amsmath}

\usepackage{mathtools}
\usepackage{amsmath}
\usepackage{amsfonts}
\DeclarePairedDelimiter{\evdel}{\langle}{\rangle}
\newcommand{\evt}{\operatorname{{\mathbb E}}\evdel}

\usepackage{stackengine}
\usepackage{url}
\usepackage{cite}
\usepackage{subfig}
\usepackage{flushend}
\usepackage{color, colortbl}
\usepackage{color,soul}
\definecolor{LRed}{rgb}{1,.8,.8}
\definecolor{LGreen}{rgb}{0.8,1,0.8}
\definecolor{HRed}{rgb}{1,.2,.2}
\definecolor{Yellow}{cmyk}{0,0,0.5,0}
\usepackage{array}
 \usepackage{booktabs} \usepackage{bigstrut}
\makeatletter
\newlength\mylena
\newlength\mylenb
\newcommand\mystrut[1][2]{    \setlength\mylena{#1\ht\@arstrutbox}    \setlength\mylenb{#1\dp\@arstrutbox}    \rule[\mylenb]{0pt}{\mylena}}
\makeatother
\usepackage{colortbl}
\usepackage{hhline}

\usepackage{bigstrut}
\newlength{\Oldarrayrulewidth}

\usepackage{tabu}

\def \MonoSense {\textit{MonoSense}}
\def \Railwaysense {\textit{Railway-Sense}}
\def \Transsense {\textit{Trans-Sense}}
\def \Rsense {\textit{R-Sense}}

\DeclareMathOperator{\atantwo}{atan2}
\IEEEoverridecommandlockouts

\begin{document}

    \setul{0.5ex}{0.3ex}
    \definecolor{Red}{rgb}{1,0.0,0.0}
    \setulcolor{Red}

\title{\Large \Transsense{}: Real Time Transportation Schedule Estimation Using Smart Phones}

\author{
\IEEEauthorblockN{Ali AbdelAziz\textsuperscript{*}}
\IEEEauthorblockA{Egypt-Japan University \\of Science and Technology\\ and Aswan University.\\
ali.abdelgalil@ejust.edu.eg }
\thanks{*Electrical Engineering department, Aswan faculty of Engineering, Aswan University, Egypt, 81542.}

\and
\IEEEauthorblockN{Amin Shoukry}
\IEEEauthorblockA{Egypt-Japan University \\of Science and Technology\\ and Alexandria University.\\
amin.shoukry@ejust.edu.eg}
\and
\IEEEauthorblockN{Walid Gomaa}
\IEEEauthorblockA{Egypt-Japan University \\of Science and Technology\\ and  Alexandria University.\\
walid.gomaa@ejust.edu.eg}
\and
\IEEEauthorblockN{Moustafa Youssef}
\IEEEauthorblockA{Alexandria University\\
Alexandria, Egypt.\\ \\
moustafa@alexu.edu.eg}}

\maketitle

\begin{abstract}
Developing countries suffer from traffic congestion, poorly planned road/rail networks, and lack of access to public transportation facilities. This context results in an increase in fuel consumption, pollution level, monetary losses, massive delays, and less productivity. On the other hand, it has a negative impact on the commuters feelings and moods. Availability of real-time transit information - by providing public transportation vehicles locations using GPS devices - helps in estimating a passenger's waiting time and addressing the above issues. However, such solution is expensive for developing countries. This paper aims at designing and implementing a crowd-sourced mobile phones-based solution to estimate the expected waiting time of a passenger in public transit systems, the prediction of the remaining time to get on/off a vehicle, and to construct a  real time public transit schedule.

 \Transsense{} has been evaluated using real data collected for over 800 hours, on a daily basis, by different Android phones, and using different light rail transit lines at different time spans. The results show that \Transsense{} can achieve an average recall and precision of 95.35\% and  90.1\%, respectively,  in  discriminating  lightrail stations. Moreover, the empirical distributions governing the different time delays affecting a passenger's total trip time enable predicting the right time of arrival of a passenger to her destination with an accuracy of 91.81\%. In addition, the system estimates the  stations’  dimensions with  an  accuracy  of 95.71\%. 


\end{abstract}
\IEEEpeerreviewmaketitle

\section{Introduction}
\graphicspath{ {./figures/} }

Using public transportation means such as light rail transit (LRT), tram, bus, or  metro is a normal daily activity  for work or leisure. However, traveling time is usually considered as a wasteful time and has a negative impact on the commuters' feelings \cite{stlouishappy2014,tyrinopoulospublic2008}. Often, passengers engage in some activities to make their transportation/commuting time more productive. Normally, public transportation follows a fixed schedule that maybe disturbed for many reasons including unpredictable traffic problems; especially in developing countries; resulting in massive delays, high fuel wastage, wasted human resources, and finally monetary and economical losses. These obstacles affect the competitiveness of public transit, which is more eco friendly than private vehicles~\cite{watkinswhere2011,elgeneidybus2009}.

A growing interest in analyzing traffic problems in developing countries has been witnessed over the years~\cite{vasconcellos2014urban,muhammad2011transportation,Roadtrafficcongestioninthedevelopingworld,strauchinvestigating2017}.  Systems like \cite{Waitingtimeperceptionsattransitstopsandstations,PassengerArrivalandWaitingTimeDistributions} are among the recent studies that attempted to estimate passengers' waiting time distributions and perceptions, depending on service and stations characteristics.
There are many definitions of the waiting time. The first expresses it as the ratio of the actual time waiting (either for getting on a vehicle  or In-Vehicle Time (IVT)) to the scheduled time~\cite{wardman_roundtable_nodate}. The second,  adopted by transportation models, assumes that average waiting times are half the service headway given random passenger arrivals \cite{Waitingtimeperceptionsattransitstopsandstations,PassengerArrivalandWaitingTimeDistributions}. The definition adopted in this paper is the first one. 
The total time spent in a trip can be estimated based on the access, regress, waiting  and IVT times~\cite{Waitingtimestrategyforpublictransportpassengers}.
Recently, many transit agencies are established to provide a suitable public transportation service and take advantage of the increasingly high-amenity transit  stations and stops to  mitigate the burden of wasted waiting time. Also, there are many facilities on social media that provide train tracking services using manual input from users. Nonetheless, these systems depend on either data available by the service providers, dedicated devices attached to the transportation vehicles, and/or manual user input. All limit their deployment, especially in develipping countries.\\

In this paper, we present \Transsense{}, a system that takes advantage of the ubiquitous devices available with the commuters, while avoiding expensive solutions based on GPS dedicated devices attached to public transit vehicles. We conduct a case study of an LRT system that takes into consideration a uniquely systematic perspective, including a wide range of stations and tram lines types, different seasons, times within the day and with many users equipped  with various mobile phones. Our results show the effectivness of the proposed technique in capturing different timing haractersitics of the trasit system dynamics.

This paper is organized as follows. Section~\ref{Sec:ProposedSystem} gives a brief introduction about the investigated tramway system as well as an overview of the proposed~\Transsense  ~system. Section~\ref{Sec:SystemComponents} describes the five main components of  \Transsense{}. 
Finally, the paper is concluded in Section~\ref{Sec:conclusion}.


	\section {Proposed System}\label{Sec:ProposedSystem}
		\subsection {Alexandria Tram System}\label{Sec:AlexandriaTramwaySystem}
	
\begin{figure}[!h]
	\centering
		\includegraphics[width=1.0\linewidth]{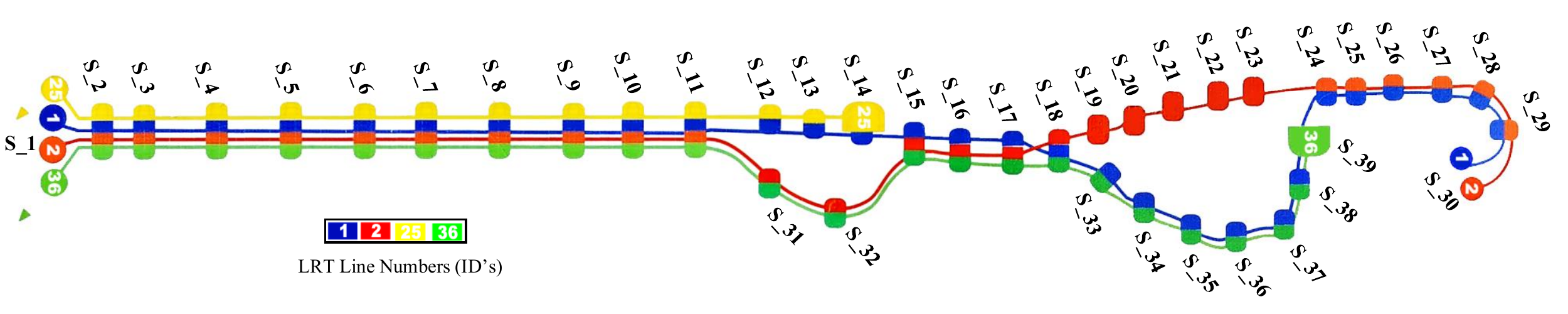}
	\caption{The investigated tram system. Two spliting areas can be identified that correlate with tram IDs.}
	\label{fig:AlexandriaTram}
	\end{figure}

Alexandria tramway started in 1860. It is the oldest in Egypt and Africa, and is among the oldest in the world. Figure \ref{fig:AlexandriaTram} and  
Table~\ref{tab:StationNamesRef} show the Ramleh Tram System, one of Alexandria LRT systems. It
runs from the east to the west of Alexandria (from Victoria to Ramleh station) and vice versa. It includes 39  stations. This system includes four different lines (four colors/Identifiers ID) that share the same rail in some parts of their journey and split into different rails in other parts.  In several locations, cars are allowed to cross the tram rails. Therefore, a tram driver has to stop until it is safe to pass or a traffic light is green. Because most traffic lines are human controlled, the waiting time at these locations is unpredictable. With the continuous increase in the population in Alexandria, migration from rural areas, the impact of the heavy traffic on trams schedule is considerable.
\begin{table}
   \caption{Different stations names and their refrences.} 
   \label{tab:StationNamesRef}
   \small 
   \centering 
   \begin{tabular}{llll} 
   \toprule[\heavyrulewidth]\toprule[\heavyrulewidth]
   \textbf{Ref.} & \textbf{Station Name}&\textbf{Ref.} & \textbf{Station Name}   \\ 
   \midrule
   S\textunderscore1 & Ramleh  &  S\textunderscore21 & Gleem \\
   S\textunderscore2 &  Al-Qa'ed Ibrahim &  S\textunderscore22 &   Fonoon Gamila\\
   S\textunderscore3 & Azarita  &  S\textunderscore23 & Zizinia \\
   S\textunderscore4 & Soter &  S\textunderscore24 &  San Stefano  \\
   S\textunderscore5 & Shoban  &  S\textunderscore25 & Thrwat  \\
   S\textunderscore6 & Shatbi &  S\textunderscore26 & Louran  \\
   S\textunderscore7 & Game3a  &  S\textunderscore27 & AlSraya \\
   S\textunderscore8 & Camb &  S\textunderscore28 &  Sidi Bishr \\
   S\textunderscore9 & Ibrahemya  &  S\textunderscore29 & El Seyoof \\
   S\textunderscore10 &  Al-Riada Al-Sughra &  S\textunderscore30 & Victoria  \\
   S\textunderscore11 &  Al-Riada Al-Kubra  &  S\textunderscore31 & Zananeri \\
   S\textunderscore12 & Kliopatra Al-Sughra &  S\textunderscore32 & SidiGaber St. \\
   S\textunderscore13 & Kliopatra Hamammat  &  S\textunderscore33 & Wezara \\
   S\textunderscore14 & SidiGaber Sheikh &  S\textunderscore34 & Felming \\
   S\textunderscore15 & Moustafa Kamel  &  S\textunderscore35 &  Bakous \\
   S\textunderscore16 & Mahfouz &  S\textunderscore36 & Safr \\
    S\textunderscore17 & Roshdi  &  S\textunderscore37 & Shods \\
   S\textunderscore18 &  Bokla &  S\textunderscore38 & Genakelese \\
   S\textunderscore19 & Hedaya  &  S\textunderscore39 & Genakelese2 \\
   S\textunderscore20 & Sapa Basha &   \\
   \bottomrule[\heavyrulewidth] 
   \end{tabular}
\end{table}


\begin{figure}
\vspace{-0.5cm}
\centering
\begin{tabular}{c}
\includegraphics[width=80mm]{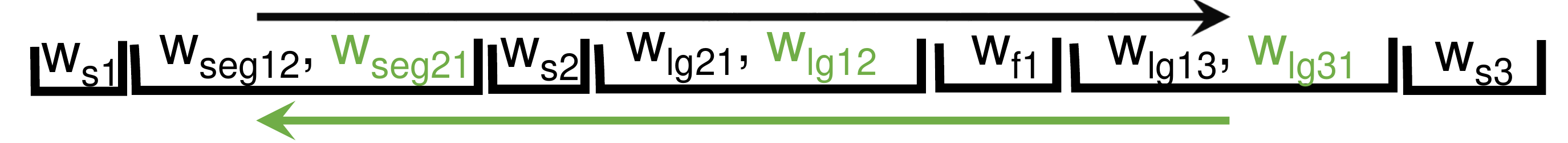}\\
   (a)  \\
\includegraphics[width=80mm]{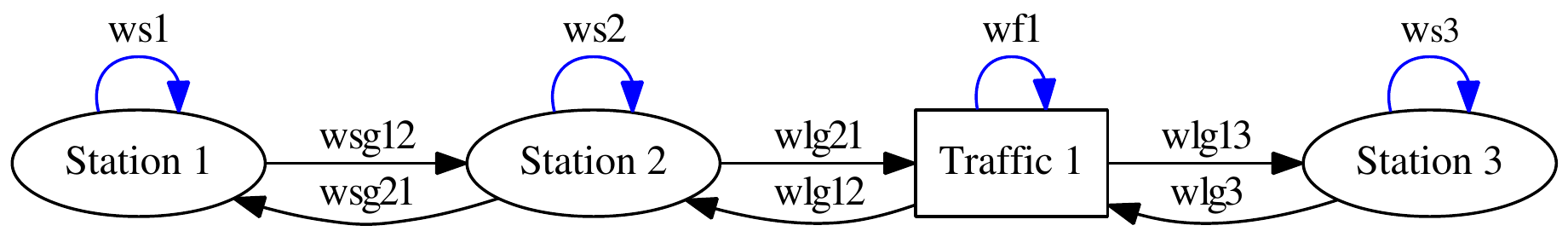}\\
(b) \\
\end{tabular}
\caption{(a) Waiting time between 3 successive stations. The route is direct between the first pair of stations while it includes a traffic light between the next pair. (b) corresponding state diagram.}
\label{DiagramWaitingtimeofStationsTripsandTrafficlight}
\vspace{-0.75cm}
\end{figure}

Whenever a tram moves from a station to the next one, one of the following scenarios is expected (Figure \ref{DiagramWaitingtimeofStationsTripsandTrafficlight}):
\begin{enumerate}
    \item The path is direct and cannot be interrupted.
    \item A cross road lies next to the starting station. An extra waiting time is required until it is safe to move. (e.g., stations S\textunderscore5 and S\textunderscore9 stations in Figure~\ref{WaitingTimesAllStation}).
    \item At least one cross-road controlled either by a human and/or traffic light interrupts the path. Hence, an extra delay time is expected.
\end{enumerate}


 The following terminology is needed in the rest of this paper:\\
 \textbf{Mandatory stop}, is a tram stop at a station.\\
 \textbf{Potential stop}, is a possible tram stop at a traffic light.\\
 All the following time dependent parameters are estimated as averages/ expectations, the operator $\evt{x}$ denotes the expectation operator.\\
\textbf{Traffic delay} $\evt{w_{f}}$, is the  average tram waiting time at a traffic signal.\\
\textbf{Station delay} $\evt{w_{s}}$, is the average tram waiting time at a station.\\
\textbf{Tram buffering delay} $\evt{w_{bf}}$, is the average tram buffering time. At peak times a tram can be waiting at a station while another tram, directly behind the first, is waiting to enter the same station. The waiting time of the second tram in this case is the buffering waiting time.\\ 
\textbf{Segment time} $\evt{w_{sg}}$, is the average time taken by a tram to travel from a given station directly to the next station (there are no cross roads or traffic lights between these two stations).\\
\textbf{Leg time} $\evt{w_{lg}}$, is the average time taken by a tram to travel from a  station/traffic signal to the next station/traffic signal on its route. This is for stations separated by traffic lights.\\
\textbf{Trip time} $\evt{w_{T}}$, is the average time for a passenger to travel from a source to a destination. It is an aggregation of multiple segments time, stations delays, traffic delays, legs time and segments time as in the following equation:
\begin{equation}\label{WaitingTimeAggregationEquation}
\begin{split}
\evt{w_{T}}=\sum_{\substack{i=1}}^{N}(\sum_{\substack{j=i+1}}^{N-1}(\sum_{\substack{k=0}}^M (\evt{w_{s_{i}}}+\evt{w_{bf_{ij}}}+\evt{w_{sg_{ij}}}\\
+\evt{w_{lg_{ik}}}+\evt{w_{f_{k}}}+\evt{w_{lg_{kj}}})))
\end{split}
\end{equation}
where, \textit{i} is the current station, \textit{j} is the next station, \textit{k} current traffic light, \textbf{N} number of stations from source to destination, \textbf{M} is the number of traffic light signals.
\subsection {Data Collection.}\label{Sec:Experiment Setup.}
The data collection took place with different mobile devices including,
Samsung Galaxy S, Samsung Note 4, HTC M9 plus, HTC E9 plus and Google Nexus. Hundred thousands of data traces  have been collected from different users at different times and seasons for about 18 months. 
 		\begin{figure*}[!h]
 	\vspace{-1cm}
	\centering
		\includegraphics[width=1\linewidth]{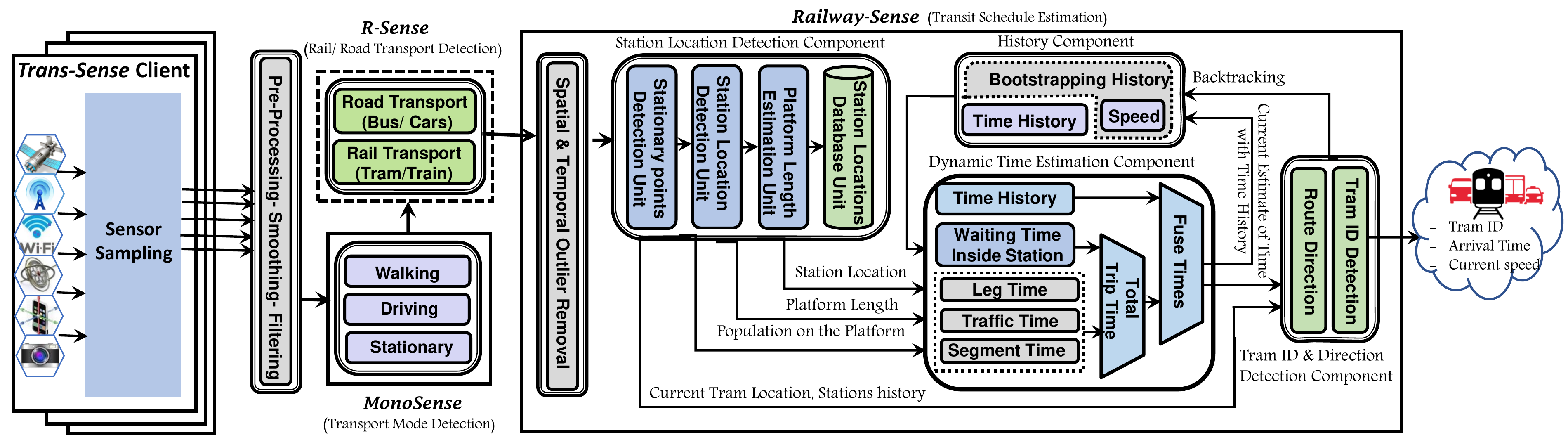}
	\caption{\Transsense{} system architecture}
	\label{fig2:Arc}
\end{figure*}

	\subsection {System Overview}\label{Sec:SystemOverview}	In this section, we present the architecture of the \Transsense{} system. It consists of three different subsystems as shown in Figure~\ref{fig2:Arc}: the \MonoSense{}, \Rsense{} and  \Railwaysense{} systems. \Transsense{} relies on crowd sensing approach to collect and accumulate data sent from the smart phones' sensors; carried by different users; to a service running on the cloud. 	The temporal and spatial information extracted from the users are correlated with their speeds and locations. The system starts by collecting a stream of GPS data ($d = d_0; d_1; \cdots;d_t;\cdots $), where each $d_t$ is an ordered pair ($Lat_t; Long_t$) representing a user latitude and longitude at time $t$. The input data stream is first filtered and smoothed, then it is passed to \MonoSense.
	
	The \MonoSense ~transportation mode detection system \cite{Monosense,Monosense2}~ differentiates between walking, riding and driving users. If the sensed data represents a moving person/ rider, it is then passed to \Rsense{}.  
	
	The \Rsense{} system identifies whether a rider is on board of a rail transportation vehicle or not. If  yes, the data is passed to \Railwaysense{}.  
	 
	The \Railwaysense{} system estimates and predicts the expected arrival times of the LRT users. A brief explanation of the \Railwaysense{} operation is given below:
	 Figure \ref{fig2:Arc} shows the  detailed components of the \Railwaysense{} sub-system. The \Rsense{} output is passed to \Railwaysense{}, then passed to the station location detection component  (Section~\ref{Sec:StationLocationDetectionComponent}) to discriminate stations' from traffic lights' stops. The Dynamic Speed Estimation component (described in Section(~\ref{Sec:DynamicTimeEstimationComponent}) answers two basic vehicle scheduling queries  based on the users' locations and views:  Station-View (\textit{SV}) (for a user waiting in a station, when will the next tram arrive)  and Vehicle-View (\textit{VV}) (for a user riding a tram, when will she arrive at her destimation). Answering these queries necessitates the estimation/ prediction of the remaining time to get on/off a vehicle.
	The history component (described in Section~\ref{Sec:HistoryComponent}) is used to aggregate previous riders' traces over time. The Tram ID, direction detection component (described in Section~\ref{Sec:TramIDandDirectionDetectionComponent}) is finally used to identify the tram direction and its ID ( that reflects its line number).
In the following section, each component of the \Railwaysense{} system architecture, is described.

\section{\Railwaysense{} System Components}\label{Sec:SystemComponents}
The five components of \Railwaysense{} are detailed in the following subsections.
	\subsection {Preprocessing Component.}\label{Sec:PreprocessingComponent}
	The main goal of this component is to reduce the  noise and remove spatial and temporal ouliars in the input raw sensors' measurements.
	\subsubsection{Temporal outliars removal}
	Temporal outliars are the outliars in the waiting time in the following parameters $\evt{w_{s}}$, $\evt{w_{f}}$, $\evt{w_{sg}}$, $\evt{w_{lg}}$  that correspond to a tram waiting at a station, a traffic delay, segment and leg times. Outliars occur due to many reasons such as accidents or tram failure delays. 
	\subsubsection{Spatial outliars removal}
\begin{figure*}[!t]
		\vspace{-0.7cm}
  \centering
  \subfloat[Original data including noise]{\label{fig:OriginalBoxPlot}\includegraphics[width=45mm]{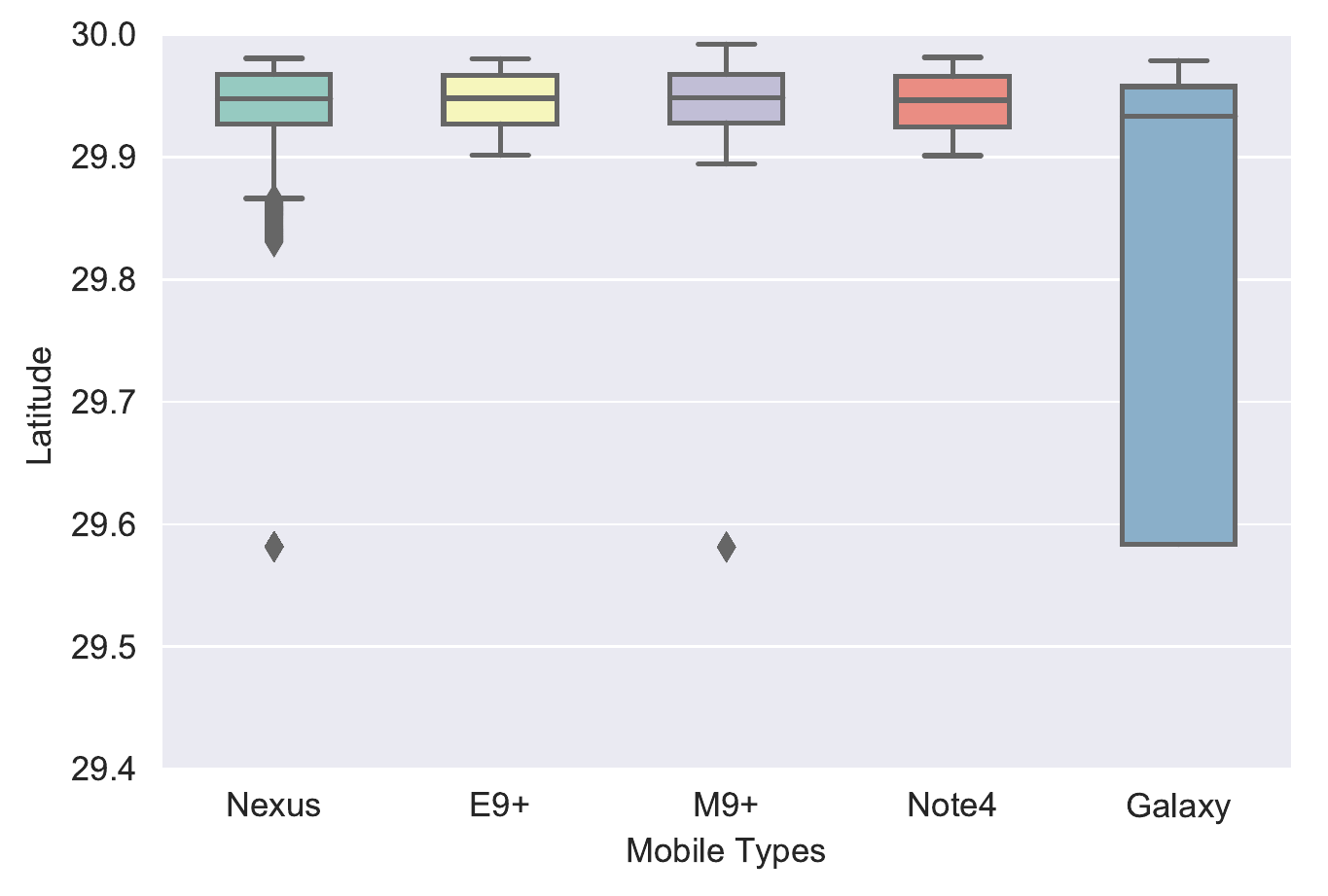}}
  \subfloat[Data after phase1]{\label{fig:Phase1BoxPlot}\includegraphics[width=45mm]{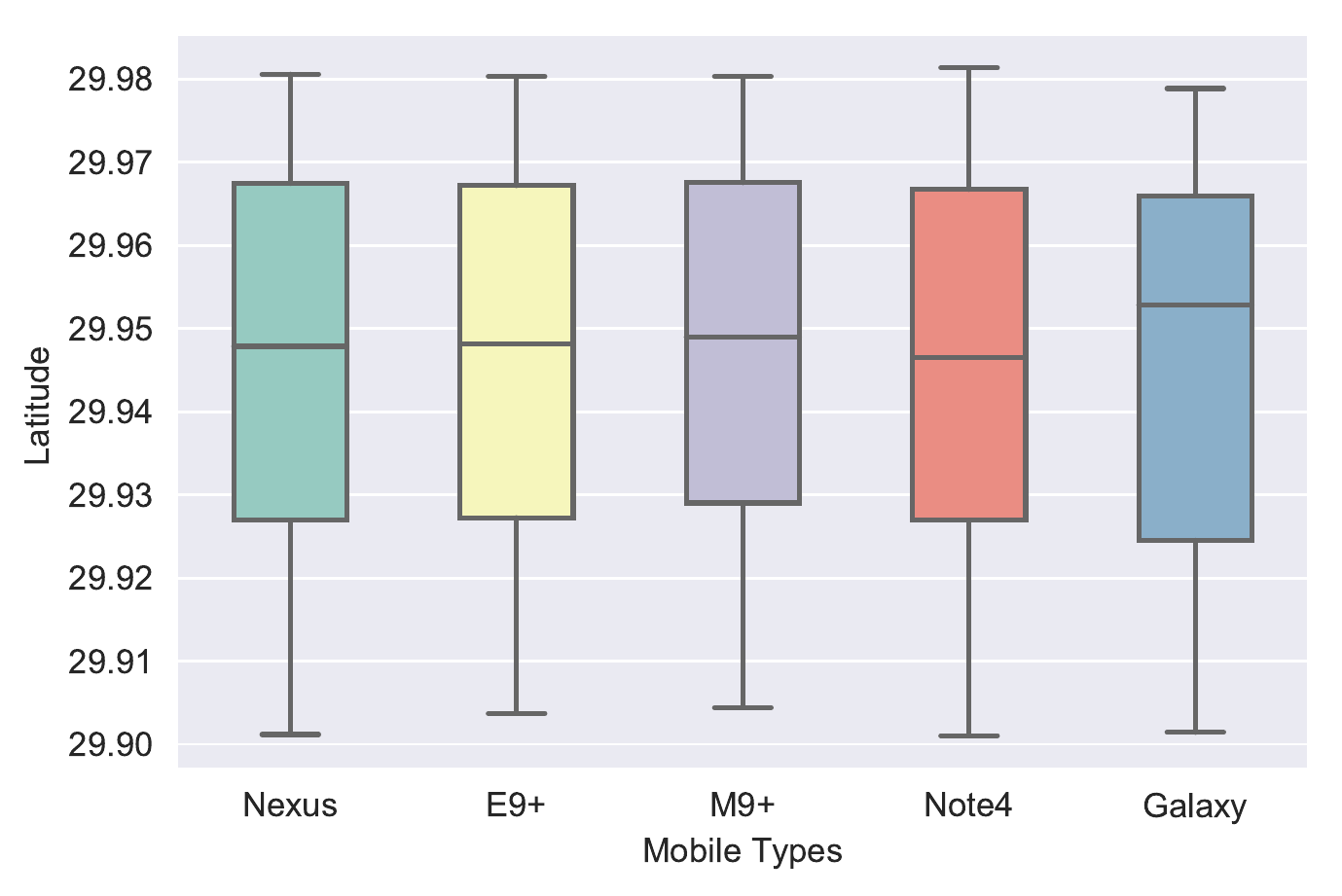}}
	\subfloat[Data after phase2]{\label{fig:Phase2BoxPlot}\includegraphics[width=45mm]{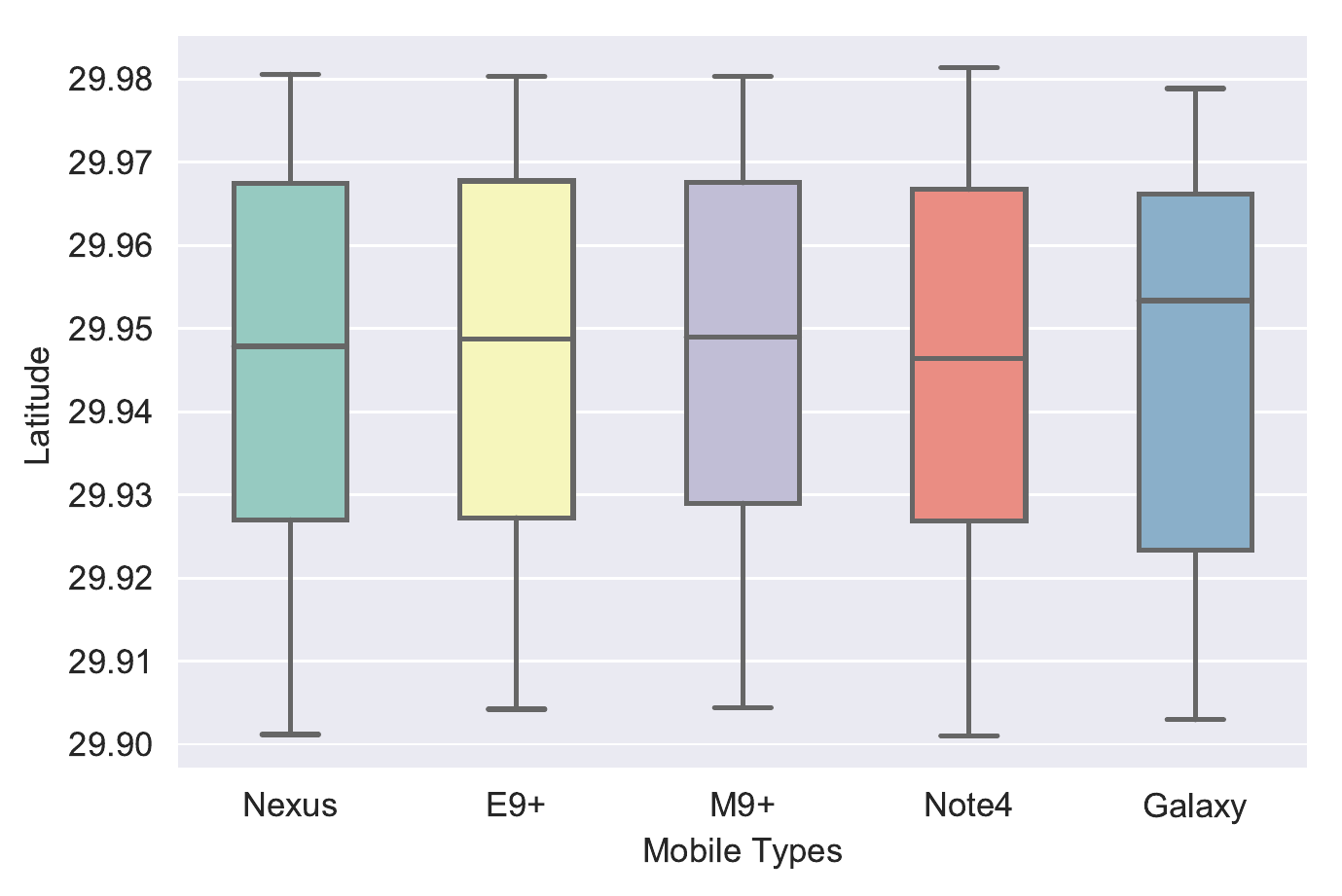}}
	\subfloat[Data after phase3]{\label{fig:Phase3BoxPlot}\includegraphics[width=45mm]{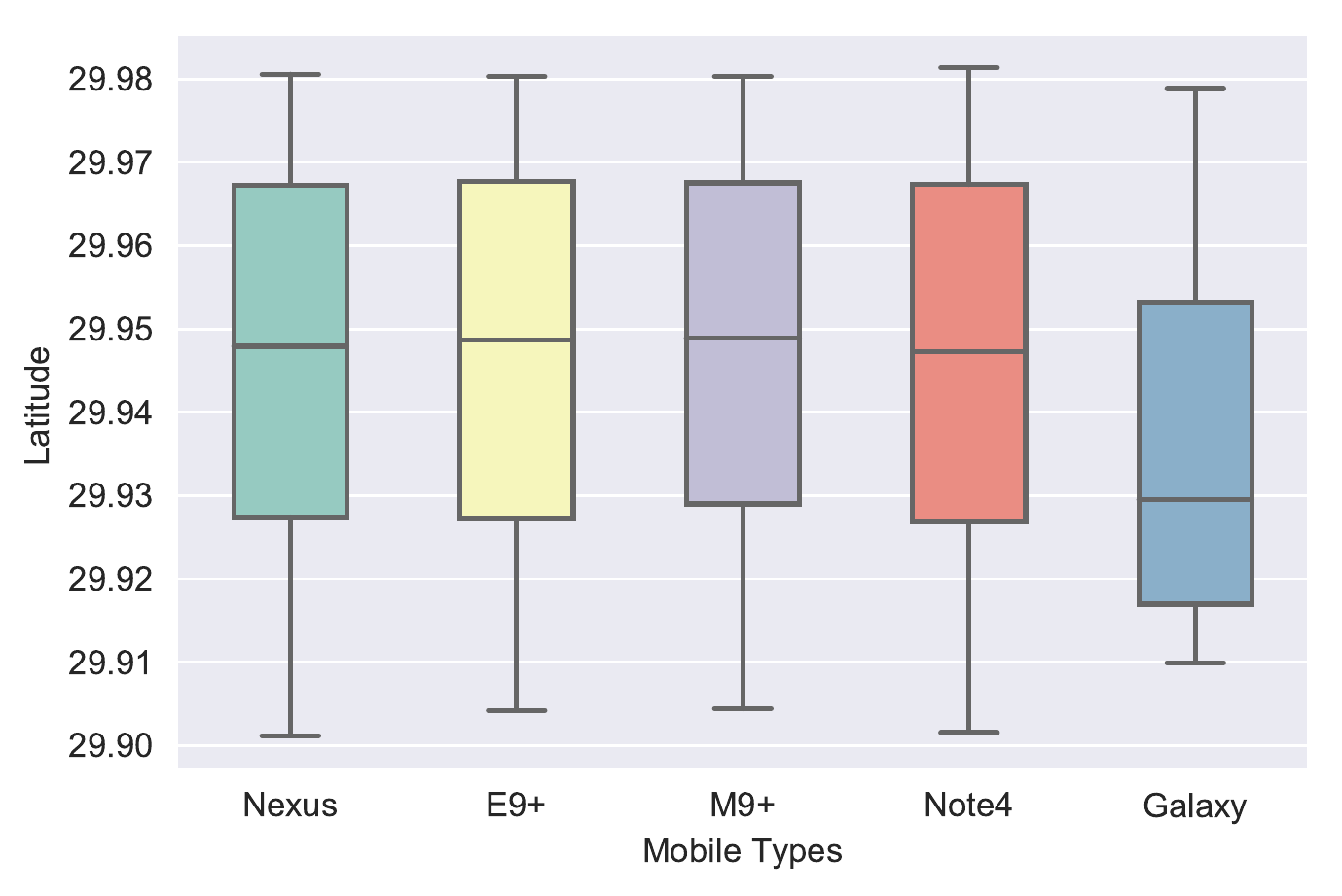}}
		\caption{Distribution of GPS data recieved from Tram users’ before and after the three preprocessing phases}
\label{fig4:BoxPlotforallDevicesWithNoise}
\end{figure*}

\begin{figure*}[htp]
 	\vspace{-0.7cm}
    \centering
\subfloat[\label{fig:OriginalE9OrangewithNoise}]{\includegraphics[width=0.25\linewidth]{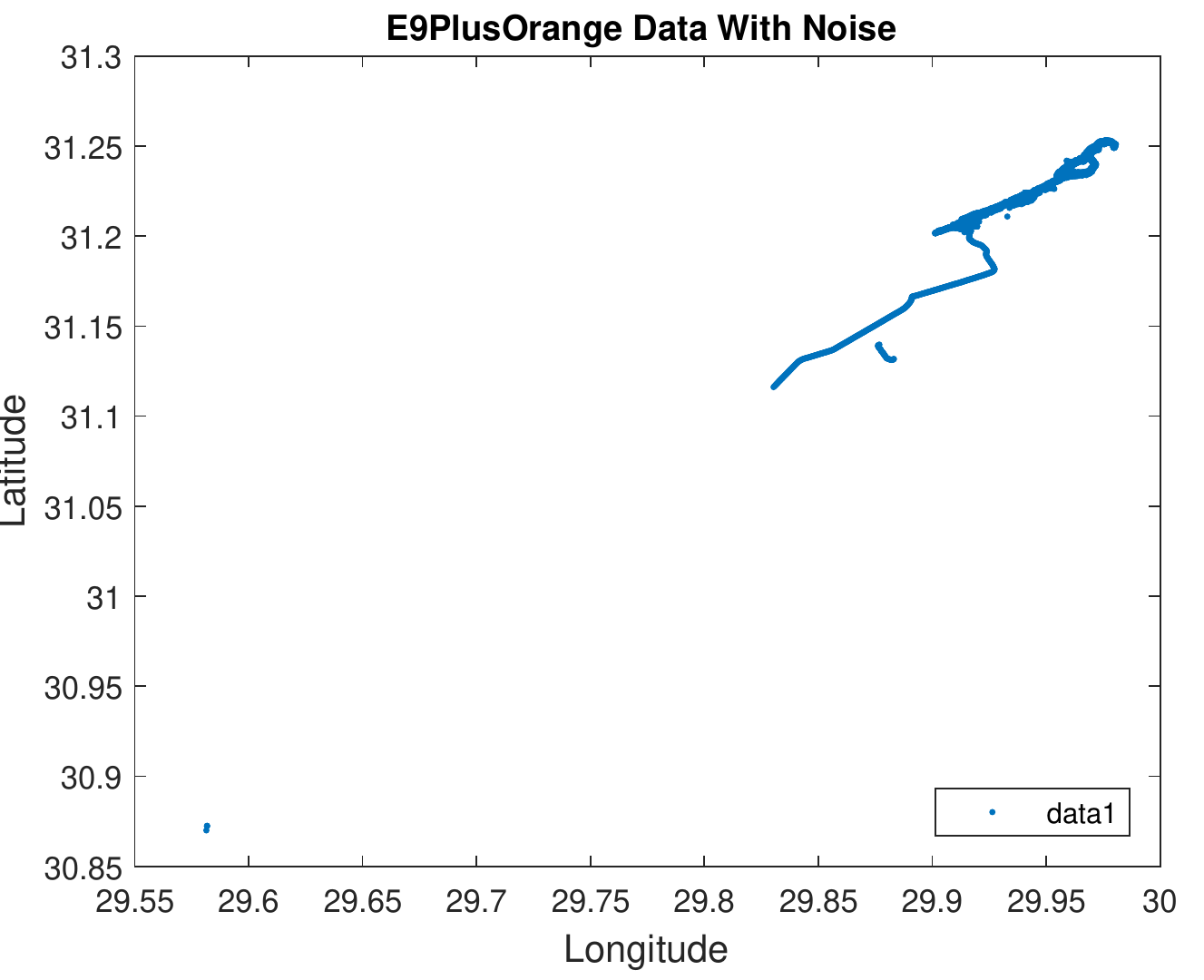}}
\subfloat[\label{fig:E9OrangeAfterPhase1SCatterPlot}]{\includegraphics[width=0.25\linewidth]{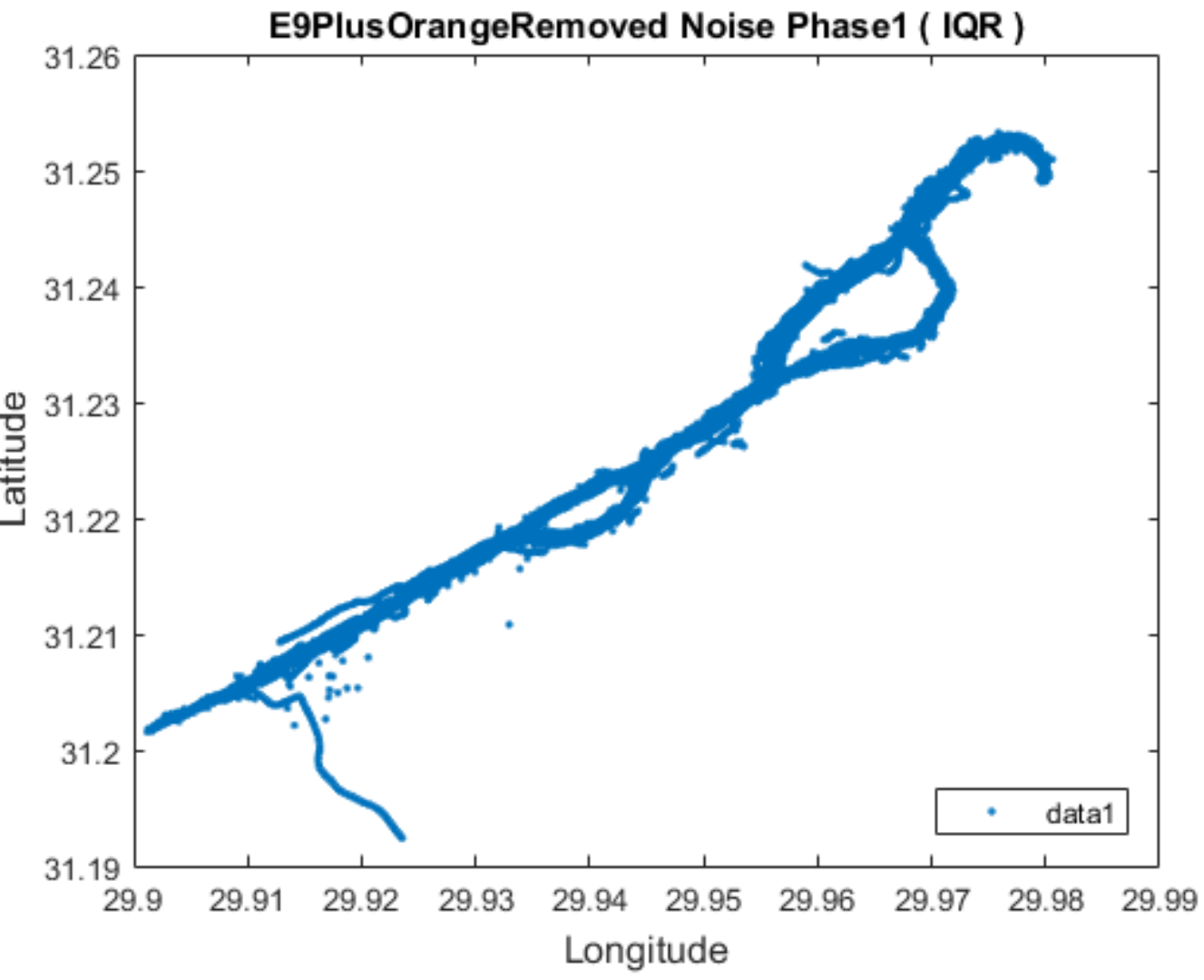}}
\subfloat[\label{fig:E9OrangeAfterPhase2SCatterPlot}]{\includegraphics[width=0.25\linewidth]{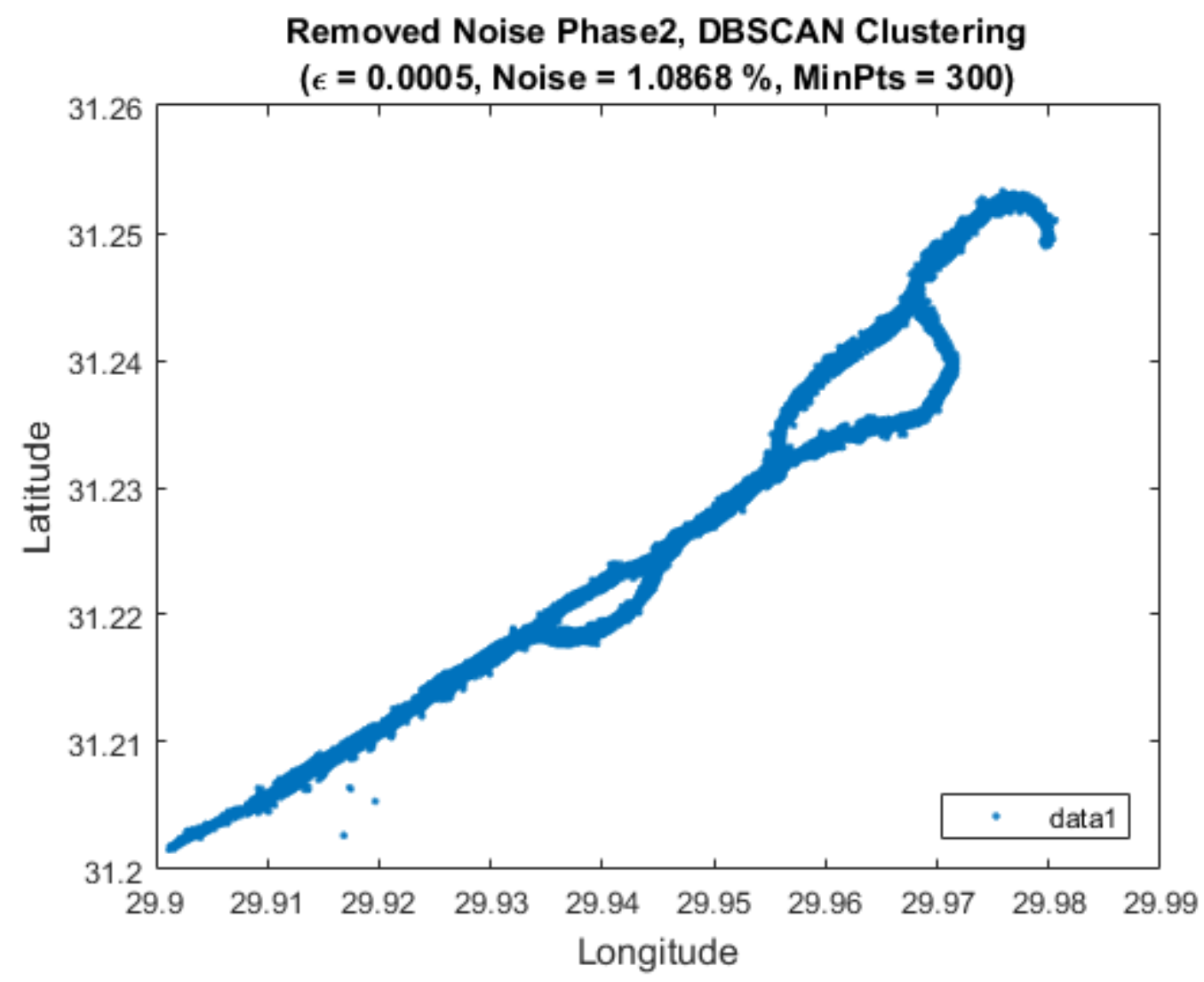}}
\subfloat[\label{fig:E9OrangeAfterPhase3SCatterPlot}]{\includegraphics[width=0.25\linewidth]{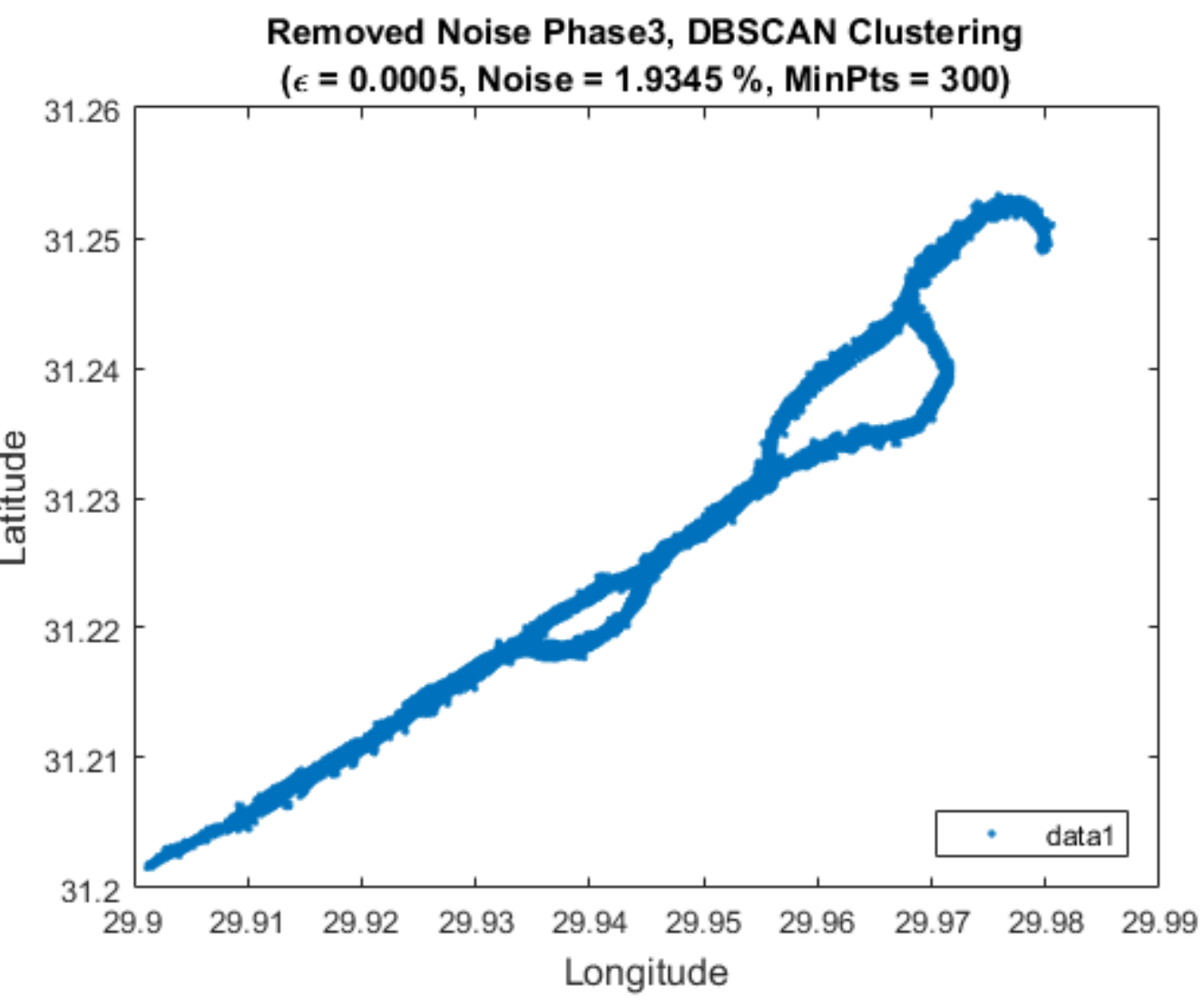}}\\
\subfloat[\label{fig:OriginalE9OrangewithNoiseHistogramPlot}] {\includegraphics[width=0.25\linewidth]{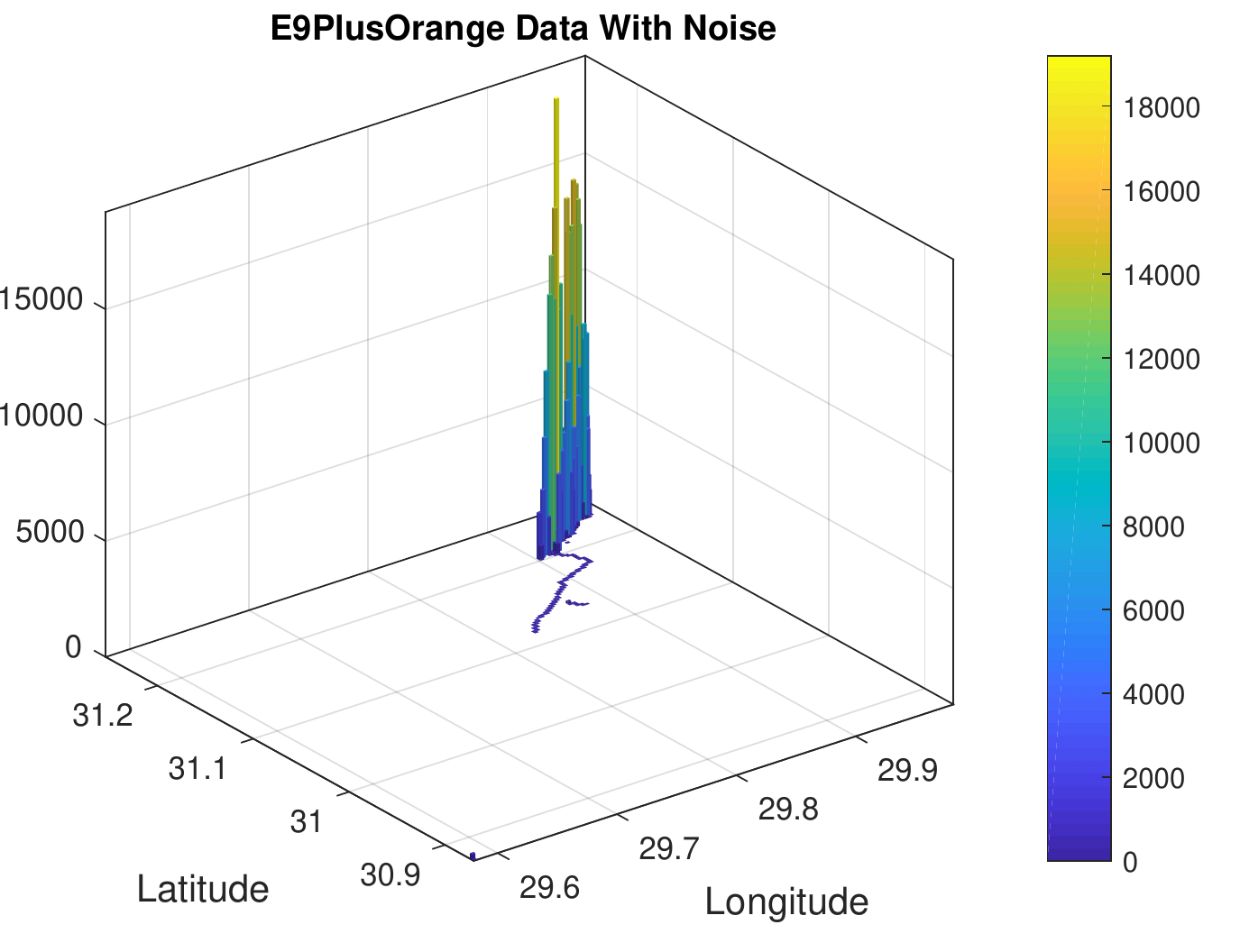}}
\hfill
\subfloat[\label{fig:E9OrangeAfterPhase1HistogramPlot}]{\includegraphics[width=0.25\linewidth]{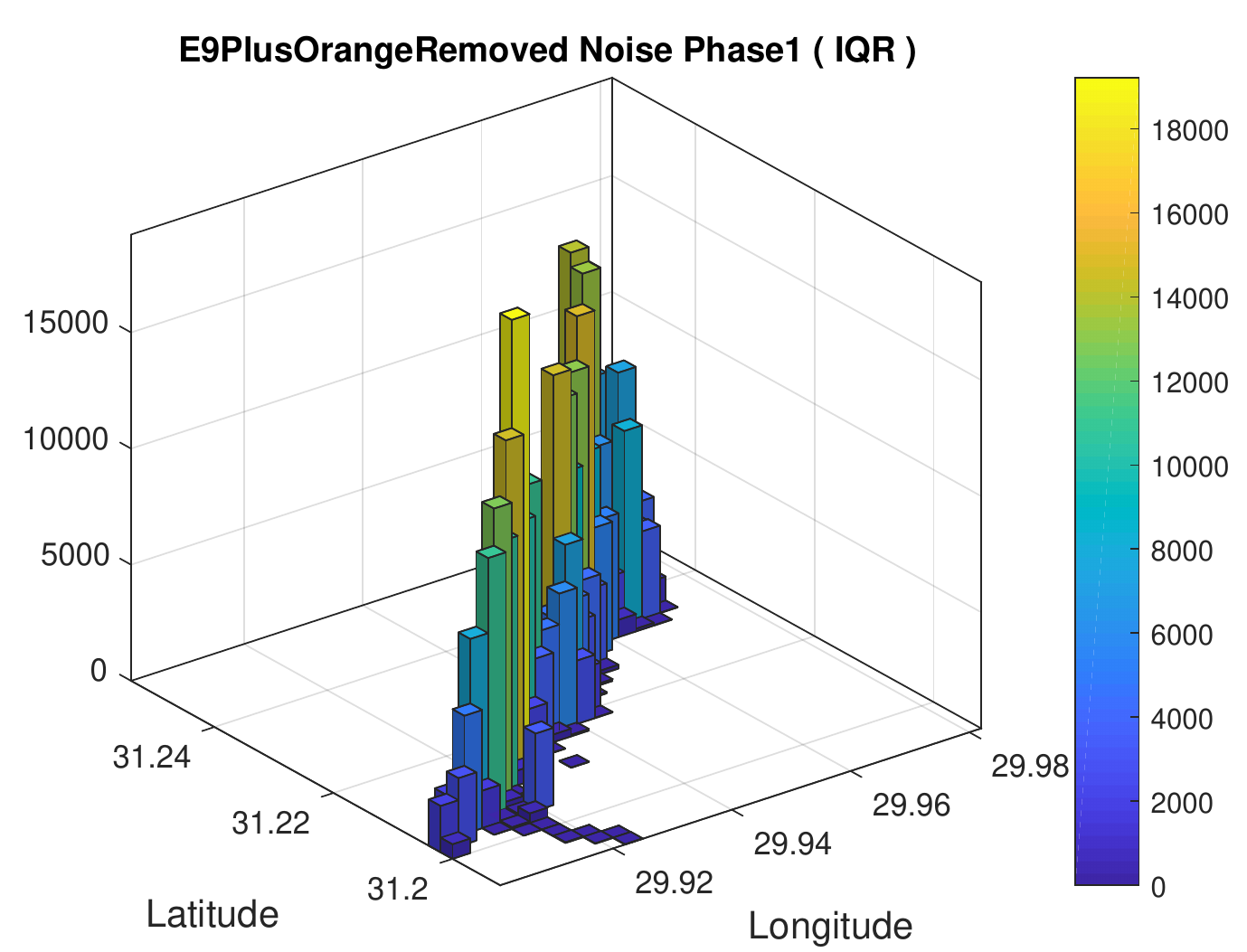}}
\hfill
\subfloat[\label{fig:E9OrangeAfterPhase2HistogramPlot}]{\includegraphics[width=0.25\linewidth]{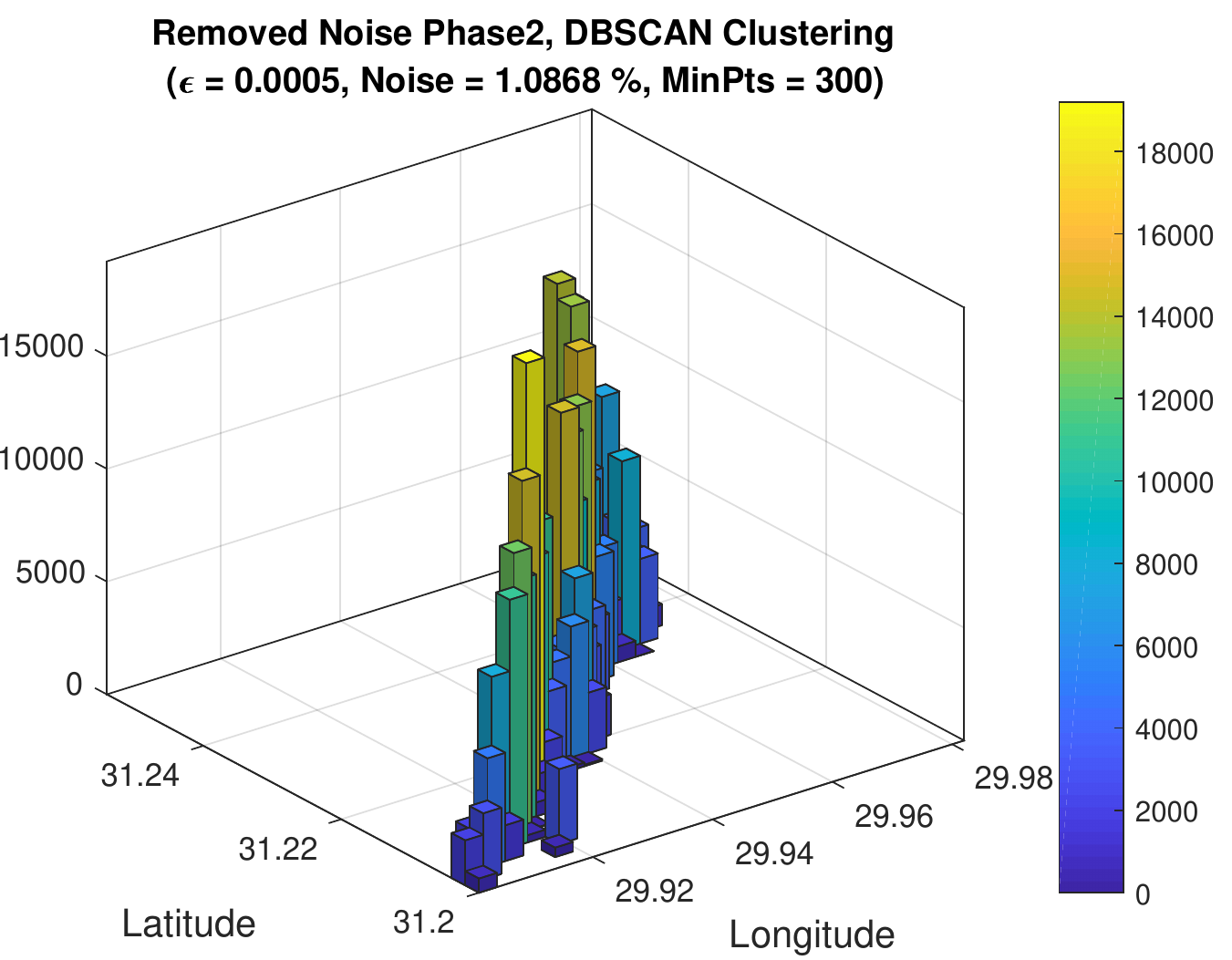}}
\hfill
\subfloat[\label{fig:E9OrangeAfterPhase3HistogramPlot}]{\includegraphics[width=0.25\linewidth]{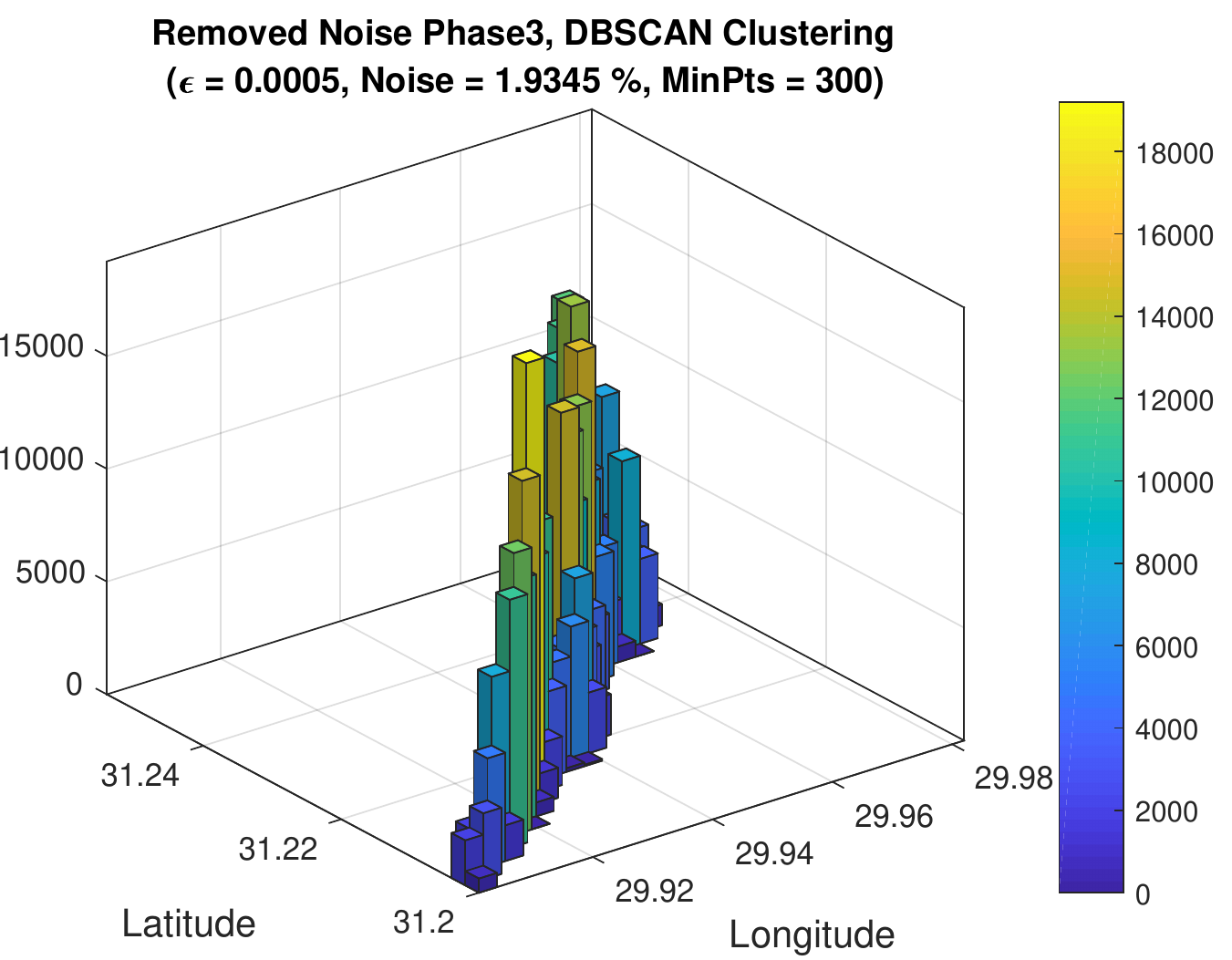}}
\caption{\small{ Data in different phases: \protect\subref{fig:OriginalE9OrangewithNoise} Original Data with noise \protect\subref{fig:E9OrangeAfterPhase1SCatterPlot} After applying Phase 1 \protect\subref{fig:E9OrangeAfterPhase2SCatterPlot} After applying Phase 2 \protect\subref{fig:E9OrangeAfterPhase3SCatterPlot} After applying Phase 3 \protect\subref{fig:OriginalE9OrangewithNoiseHistogramPlot}, \protect\subref{fig:E9OrangeAfterPhase1HistogramPlot}  \protect\subref{fig:E9OrangeAfterPhase2HistogramPlot} \protect\subref{fig:E9OrangeAfterPhase3HistogramPlot} Histograms of original data, after Phase 1, after Phase 2 and after Phase 3, respectively. }}  
\label{fig:E9OrangeOriginalDataWithNoise}
    \end{figure*}

Spatial data contains noise and outliars mainly due to the urban canyon effect and or/ inaccurate position information \textit{(e.g. glitches)} \cite{mao1999noise,aly2013dejavu}. Specifically, errors are more often when the GPS signal bounces off \cite{giremus2007particle,kos2010effects,chansarkar2003resolving}; where a user reading changes vastly back and forth among different nearby areas as shown in Figures \ref{fig4:BoxPlotforallDevicesWithNoise}, \ref{fig:E9OrangeOriginalDataWithNoise}.

	Moreover, misleading readings are due to \textit{'multipath effect'} \cite{townsend1994practical,van1994multipath,soubielle2002gps}, where the direct path to the GPS receiver is blocked. For the above reasons, spikes and spurious changes occur and lead to substantial deviations in the measured locations. To handle these errors, we apply three different filtering phases: coarse outliar removal (Phase 1), noise filtering (Phase 2), and duplicates removal (Phase 3). The coarse filtering removes outliar data that is significantly different than other \textit{(unusual identifiable values)}. It uses the empirical three sigma rule of thumb\footnote{It states that for a  normal distribution 99.7\%  of the data lie within a range of three standard deviations of the \textit{mean}, [$\mu$-3$\sigma$, $\mu$+3$\sigma$]} \cite{wheeler1992understanding,Akbari2016,black2011business} as shown in Figure~ \ref{fig:E9OrangeOriginalDataWithNoise}.

The filtering phase uses a smoothing filter with non-overlapping windows to smooth the data  and clusters similar data points into groups using the density based spatial clustering algorithm (DBSCAN) \cite{ester1996density}. \textbf{\textit{Dbscan}} is a well-known clustering algorithm that requires only two parameters not including the number of clusters:
		\begin{itemize}
			\item \textbf{\textit{MinPts}}: specifies the minimum number of points in the neighborhood of a given point in order to be included in a cluster.
			
       \item \textbf{\textit{Epsilon $\epsilon$}}:  specifies the size (radius) of the (circular) neighborhood.
		\end{itemize}
Finally, dissimilar remaining points are removed. Among the obtained clusters, there are clusters that have multiple data points at the same location with lower variance. They represent duplicate entries that are removed as shown in Figure \ref{fig:E9OrangeOriginalDataWithNoise} (b,d).

The filtered data (98.65\% of the raw data) is then passed to the station location detection component to discriminate between stations and traffic lights according to passengers' behavior.

	\subsection { Stations Locations Detection Component.}\label{Sec:StationLocationDetectionComponent}
	The main function of this component is to extract the semantics hidden in the filtered GPS points. Since a station occupies a relatively large area, it corresponds to a large number of GPS measurements. GPS measurements, at the same physical location, can vary by meters either due to GPS characteristics \cite{soubielle2002gps} or receiving capabilities (differences in the GPS receiver accuracy among the sensing devices).  
	Therefore, we cluster these data into meaningful places. These places correspond to either stations (at which  passengers are waiting  for a tram ) or moving/ stationary trams. Stationary trams are those waiting at stations or traffic lights.
	In order to discriminate between these two tram states, we make use of the travel environment as well as passengers’  behavior.

	\subsubsection { Stationary Points Detection Unit}\label{StationaryPointsDetectionUnit}
\begin{figure}
\hspace{-0.5cm}
\centering
\includegraphics[width=80mm]{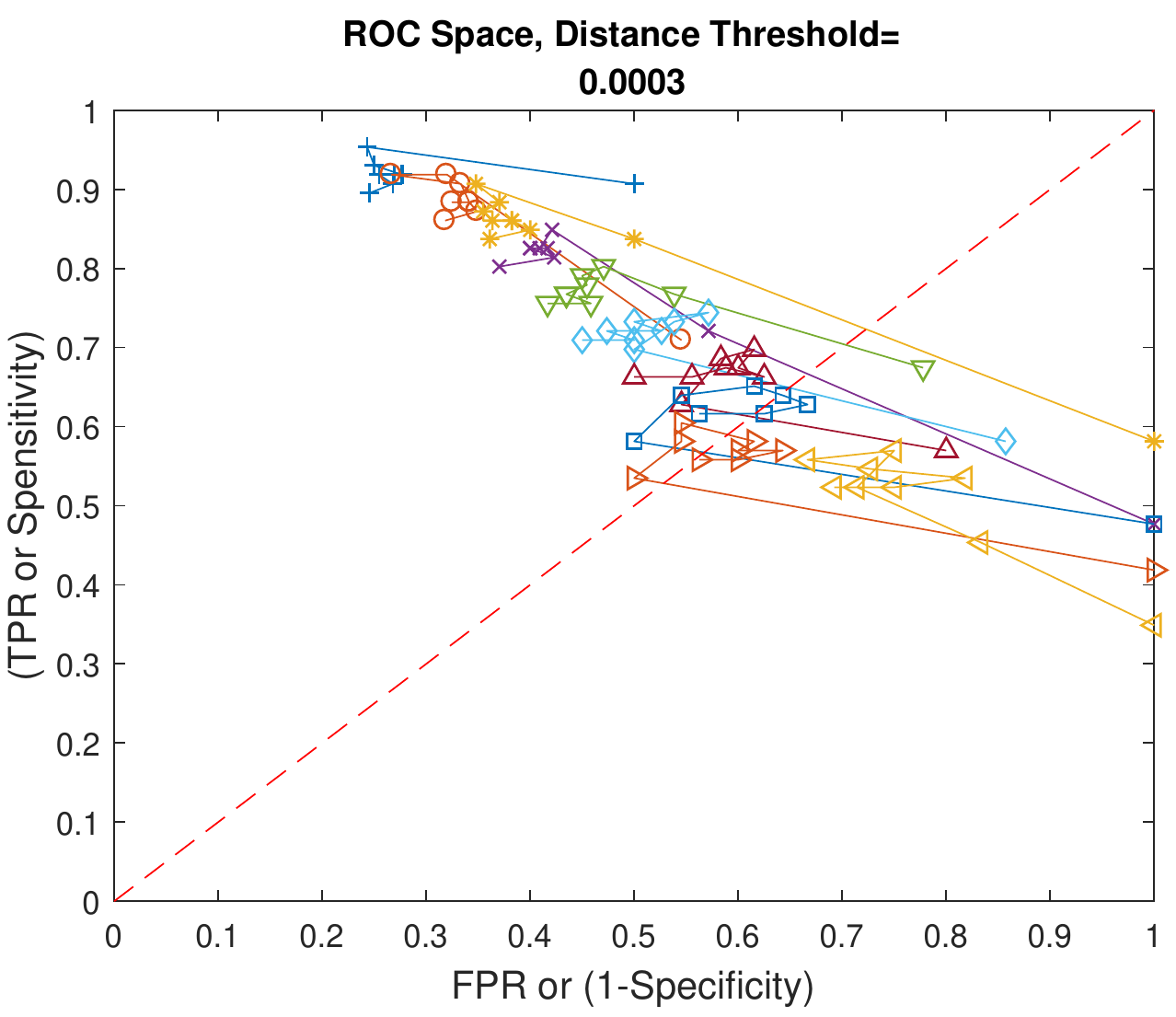}
\caption{Optimal Clustering Parameters}
\label{OptimalClusteringParameters}
\end{figure}

	The main purpose of this unit is the automatic detection of the stationary points (\textit{detection of  passengers' inside a stand still tram that is waiting at a  station or traffic light}). The unit starts by the stream of preprocessed random points. These points are clustered into stationary or moving points. Points that have nearly zero speed are considered as stationary points. Further, stationary points are  clustered into either stations or traffic lights. We have used Dbscan to detect the stationary points. 	
The optimal parameters \textit{Minpts*} and \textit{Epsilon*} for the collected data have been found based on ground truth data (the true locations of the tram stations and traffic lights) and repetitive application of the Dbscan algorithm with a range of values for \textit{Minpts} and \textit{Epsilon}. The centroids of the found clusters are calculated and matched to the true centroids. If a matching is less than a small distance threshold (\textit{DT}), it is considered as a positive hit. Figure~\ref{OptimalClusteringParameters} illustrates several ROC curves corresponding to different trials of Dbscan algorithm (and different values of \textit{Minpts, Epsilon, DT}). From Figure~\ref{OptimalClusteringParameters}, the optimal parameters \textit{Minpts*, Epsilon*} and \textit{DT*} that achieve maximum TPR (\textit{True Positive Rate})= \textit{0.9535} and minimum FPR (\textit{False Positive Rate})= \textit{0.2432} and \textit{DT= 0.0003} have been recorded and listed in Table~\ref{ClusteringParameters}.

		
	\begin{table}[htbp!]
	\centering
	\begin{tabular}{*8l}    \toprule
\emph{\textbf{Parameter/Metric}} & & & &\emph{\textbf{Value}} \\\midrule
MinPts    & & & &100   \\ 
Epsilon $\epsilon$   & & & &\ang{0.0002;;}  \\ 
Distance Threshold DT   & & & &\ang{0.0003;;}    \\ 
TPR (True Positive Rate)  &  & & &0.9535   \\ 
FPR (False Positive Rate)   & & & &0.2432   \\ 
\rowcolor{green!50} Precision& & & &90.1\% \\ 
\rowcolor{green!50} Recall & & & &95.35\% \\\bottomrule
 \hline
\end{tabular}	
\begin{tablenotes}
     \item[1]\hspace{1.5cm}Note that $\epsilon$, $D_{t}$ are in degrees:\par
		\setlength\parindent{25pt} \hspace{1.5cm}\ang{1;;} $\approx$ 111 km (110.57 eq’l — 111.70 polar).\par
      \hspace{1.5cm}\ang{;1;} $\approx$ 1.85 km (= 1 nm)	\ang{0.01;;} $\approx$ 1.11 km.\par
      \hspace{1.5cm}\ang{;;1}	$\approx$ 30.9 m \ang{0.0001;;} $\approx$ 11.1 m.\\ .
   \end{tablenotes}
		\caption{Optimal clustering parameters and corresponding performance metrics}
	\label{ClusteringParameters}
\end{table}	

		\subsubsection { Station Location Detection Unit.}\label{Station Location Detection Unit}
The function of this component is to discriminate between stations and traffic lights based on either temporal or  spatial  methods. From the temporal point of view, the real data shows that, in general, the magnitude of $\evt{w_{s}}$ is greater than $\evt{w_{f}}$. Figure \ref{WaitingTimesAllStation} shows stations' and traffic lights' samples waiting times ($\evt{w_{s}}$, $\evt{w_{f}}$) of trams in one of the tram lines. At \textbf{\textit{S\textunderscore5 station}} one may note that the waiting times are, in general, longer than other stations because all types of waiting times ($\evt{w_{s}}$, $\evt{w_{f}}$ and $\evt{w_{bf}}$) usually occur at this station.\\
From the spatial point of view,  both stations and traffic lights are places of frequent stops. However, the frequency of stops at stations is considerably larger than the stops at traffic lights. Moreover, stations and traffic lights can be differentiated based on the passengers in/out-flow patterns to/from the trams, at these locations. Commuters engage in different activities while getting on/off trams such as moving up/down, standing, sitting, turning, etc... These activities can discriminate between traffic lights and stations. Since travelers are expected to get off/ on (from/ to) a tram only at stations while remaining relatively motionless at traffic lights, the corresponding passengers' activity patterns can discriminate between these two situations. These activity patterns ~\cite{CrowdMeter,ElhamsharyActivityRecognitionofRailway} (illustrated in Table \ref{tab:GettingOnOffPassengersActivityStates}) can be recognized using either dedicated wearable sensors or the ubiquitous sensors in smart phones via pervasive computing\cite{LandmarkSense}.
\begin{table}[htbp!]
	\begin{minipage}{4cm}
\begin{threeparttable}[t]
	\centering
\begin{tabular}{|c|c|} 
 \hline
\textbf{Activity} & \textbf{Activity Description}\\\hline
 Getting on a Tram \tnote{1}& \multicolumn{1}{m{6cm}|}{Standing, walking, turning Left or right, walking, climbing up stairs, turning left or right, walking again till reaching seat, turning left or right 90$^{\circ}$ or 180$^{\circ}$, seating down, sitting for long time.}  \\\hline 
 Getting off a Tram\tnote{2}& \multicolumn{1}{m{6cm}|}{Sitting, standing up, turning left or right 90$^{\circ}$ or 180$^{\circ}$, walking till reaching the door, turning left or right to be in front of the stairs, getting down, walking on the ground.}  \\ 
 \hline
\end{tabular}
\begin{tablenotes}
     \item[1] Note that \textit{Getting On Sequence} will start from steady standing in the station.
     \item[2] Note that \textit{Getting Off Sequence} will start from steady sitting state in the Tram.
   \end{tablenotes}
\end{threeparttable}
\end{minipage}
	\caption{Getting on/off passengers' activity patterns.}
	\label{tab:GettingOnOffPassengersActivityStates}
\end{table}
\subsubsection { Platform Length Estimation Unit.}
\begin{figure}
\hspace{0cm}\includegraphics[width=80mm]{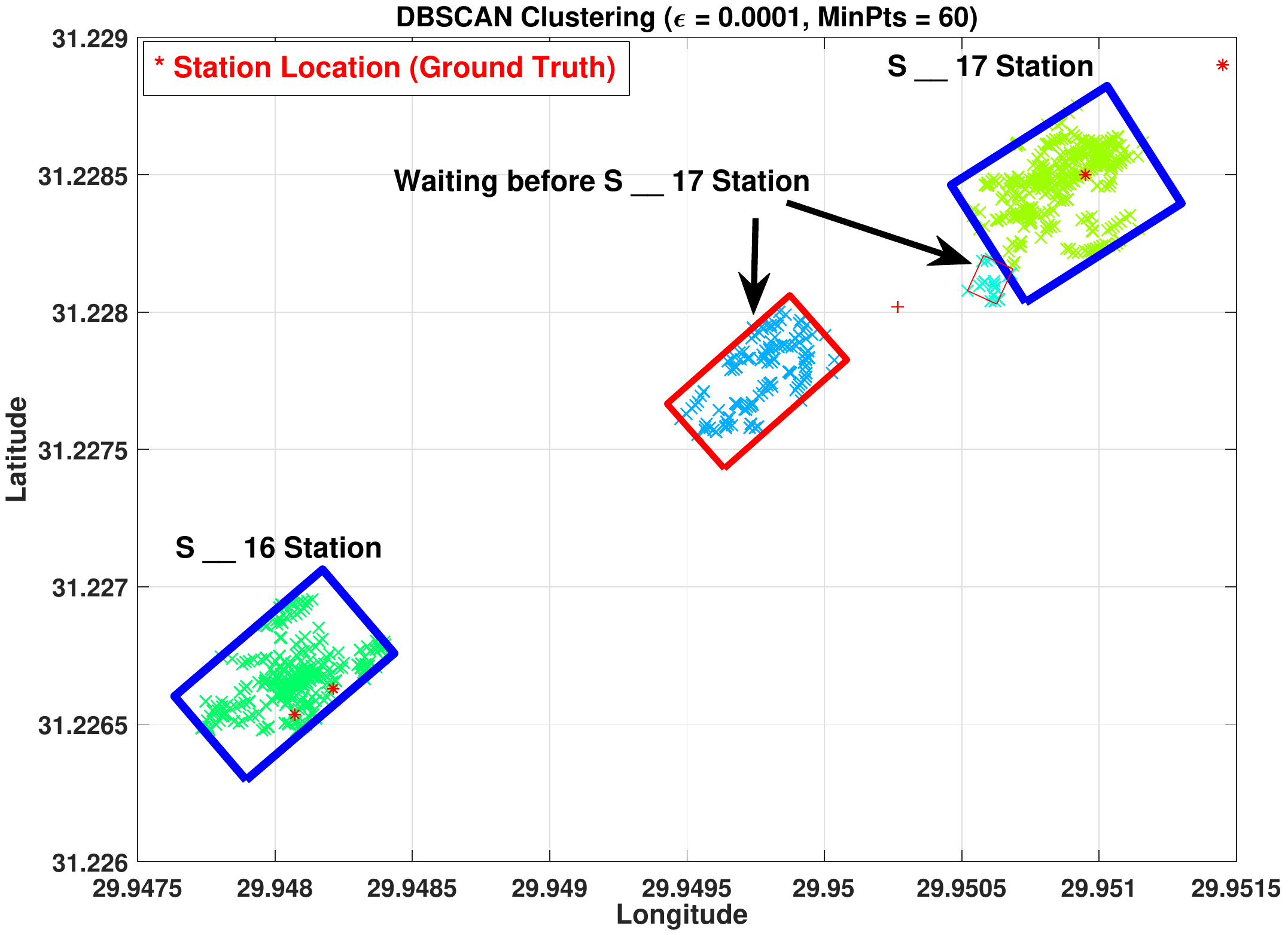}
\caption{ The red box denotes a tram waiting at a traffic light between S\textunderscore16 and S\textunderscore17 stations (blue boxes)}
\label{BoundingRectangle}
\end{figure}

The function of this unit is to estimate the platform length after splitting stations from traffic lights using the previous unit. Platform length detection identifies whether the tram reached a station or just buffering outside the station, as shown in Figure \ref{BoundingRectangle}. 
The minimum bounding rectangle (\textbf{\textit{MBR}}) of the clustered points representing a station, is used to estimate the platform length. The length of the larger side of the bounding rectangle represents the platform length. Given that ($Long_{1}, Lat_{1}$) and ($Long_{2},Lat_{2}$) are the coordinates of the end points of the larger side,  Eq  \eqref{Equations} shows how to calculate a station platform length(D) using Haversine Formula: \cite{noauthor_nasa_nodate}.
\begin{equation} \label{Equations}
\begin{split}
(D_{long}, D_{lat}) =& (Long_{2}, Lat_{2}) - (Long_{1}, Lat_{1})\\
A =& \cos(Lat_{1}) * \cos(Lat_{2}) * (\sin(\frac{D_{long}}{2}))^2 \\
& + (\sin(\frac{D_{lat}}{2}))^2  \\
C =& 2 * \atantwo( \sqrt{A}, \sqrt{(1-A)} ) \\
D =& R * C (\emph{where R is the radius of the Earth})
\end{split}
\end{equation}
\footnotetext{This formula does not take into consideration the (ellipsoidal) shape of the Earth. It will tend to overestimate trans-polar distances and underestimate trans-equatorial distances. The values used for the radius of the Earth $(3961 miles, 6373 km)$ are optimized for locations around 39 degrees from the equator (roughly the Latitude of Washington, DC, USA).}
We could correctly estimate stations' platforms lengths with an accuracy of 95.7\%. For example, the true platform length of S\textunderscore17 station is 70 meters, while the estimated length is 67 meters. As a result, we can integrate new stations to Google maps and other similar services.
		
	\subsubsection { Stations locations database unit.}
	The function of this unit is to store stations' locations (Long, Lat) and dimensions and add new stations directly after finding them from previous components.
  \subsection {Dynamic Time Estimation Component.}\label{Sec:DynamicTimeEstimationComponent}
\begin{figure}
\includegraphics[width=80mm]{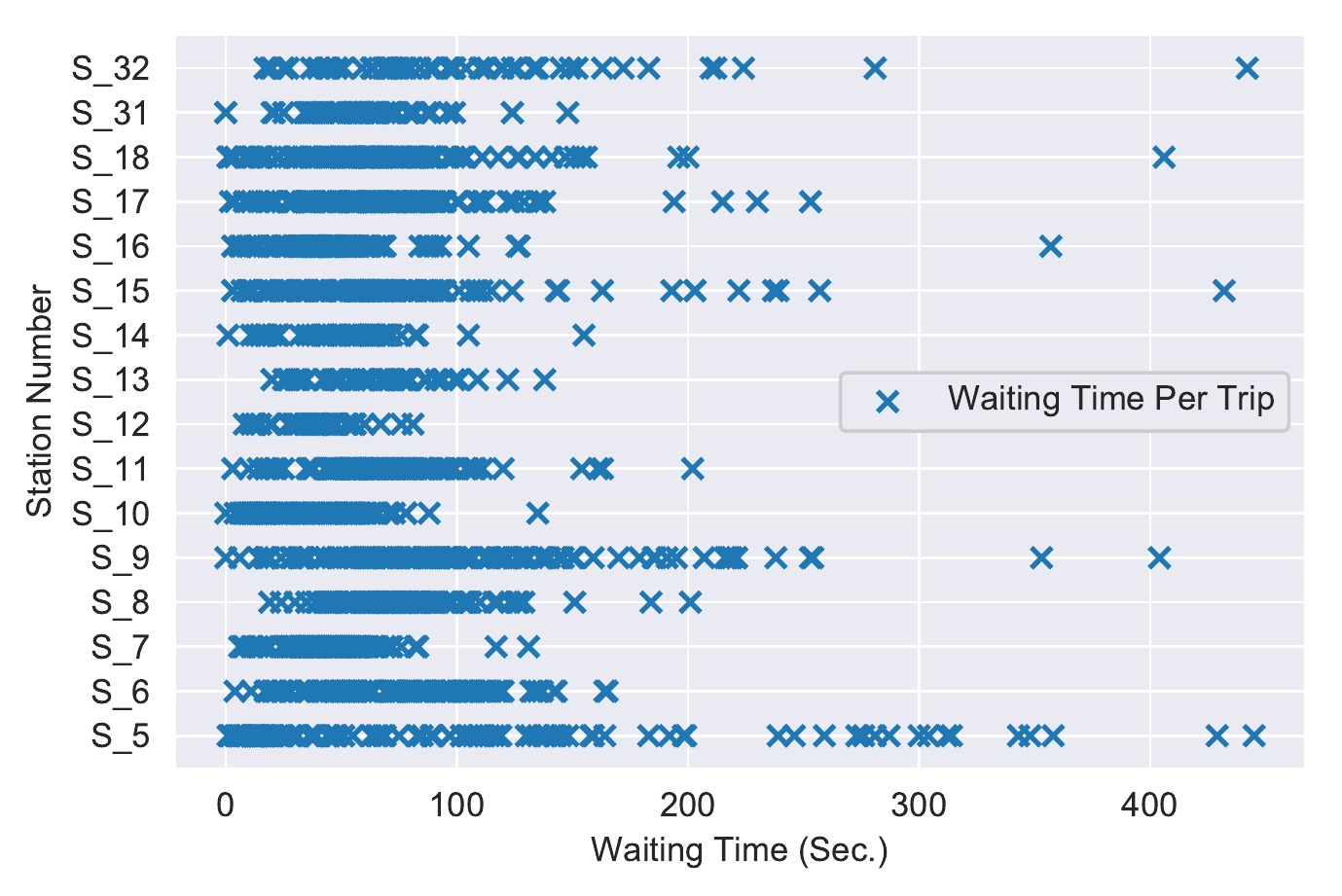}
\caption{ Stations' and traffic lights' samples waiting times ($\evt{w_{s}}$, $\evt{w_{f}}$) of trams in one of the tram lines.}
\label{WaitingTimesAllStation}
	 \vspace{-0.8cm}
\end{figure}

  This component estimates the expectation of each of the random parameters defined in Eq~\ref{WaitingTimeAggregationEquation}. Towards this goal the distribution governing each parameter is plotted and its expectation is calculated. For example, Figure~\ref{WaitingTimesAllStation} plots samples of the waiting time ($\evt{w_{s}}$) for different stations for many trips.  A GPS sample observation is identified as belonging to a passenger in a tram waiting at a station if the spatial coordinates lie within the station boundaries (identified in the previous sub-section). The time stamp of this observation belongs to the set of time stamps that define the tram waiting time interval.
  In some stations in Figure~\ref{WaitingTimesAllStation}, such as S\textunderscore5 station, the waiting time is long because it is  directly followed by a traffic light. Based on real data,  it ranges from 30 to 400 seconds.

\begin{figure}
	\vspace{-0.7cm}
\begin{tabular}{cc}
\hspace{0cm}\includegraphics[width=40mm]{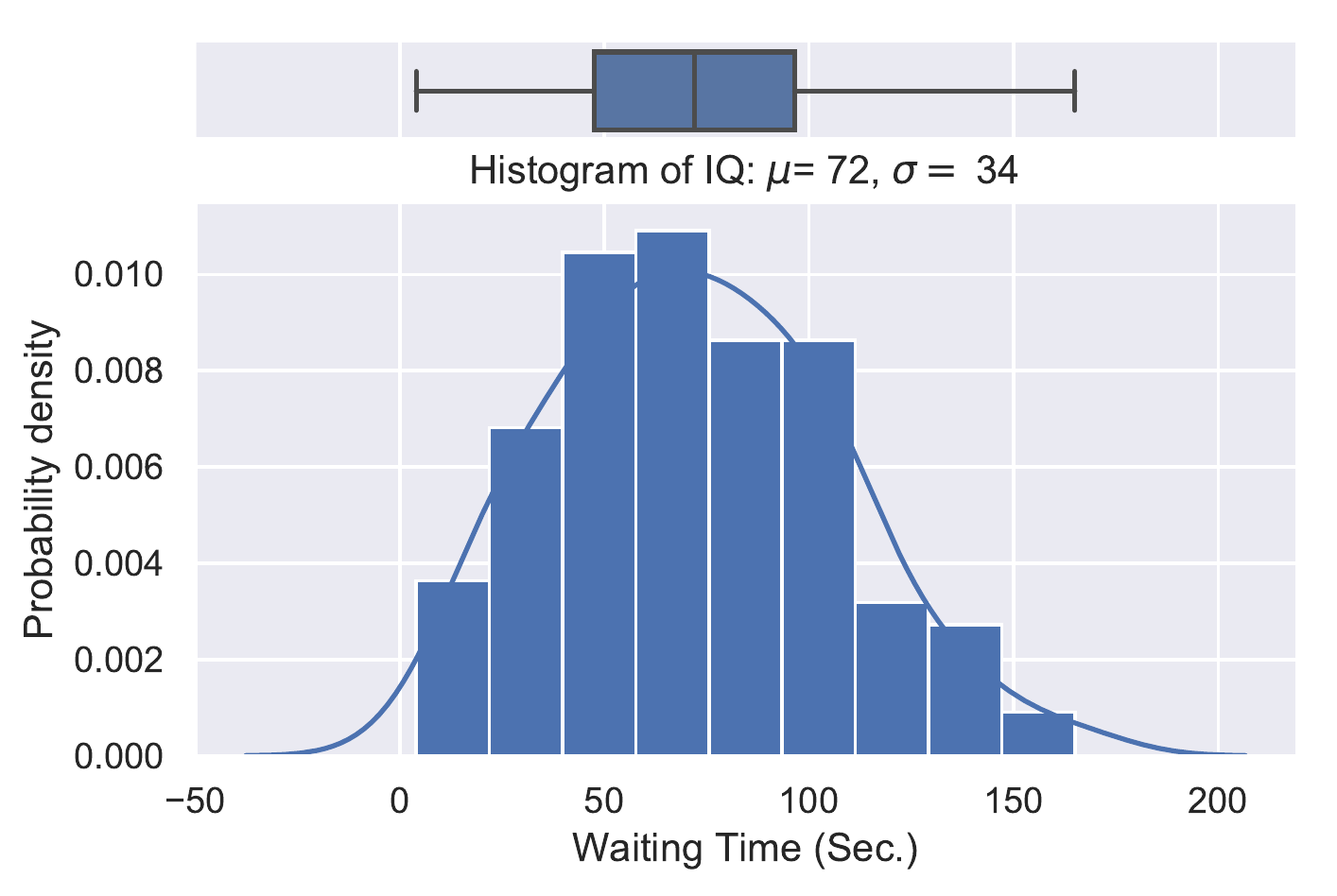} &
\hspace{0cm}\includegraphics[width=40mm]{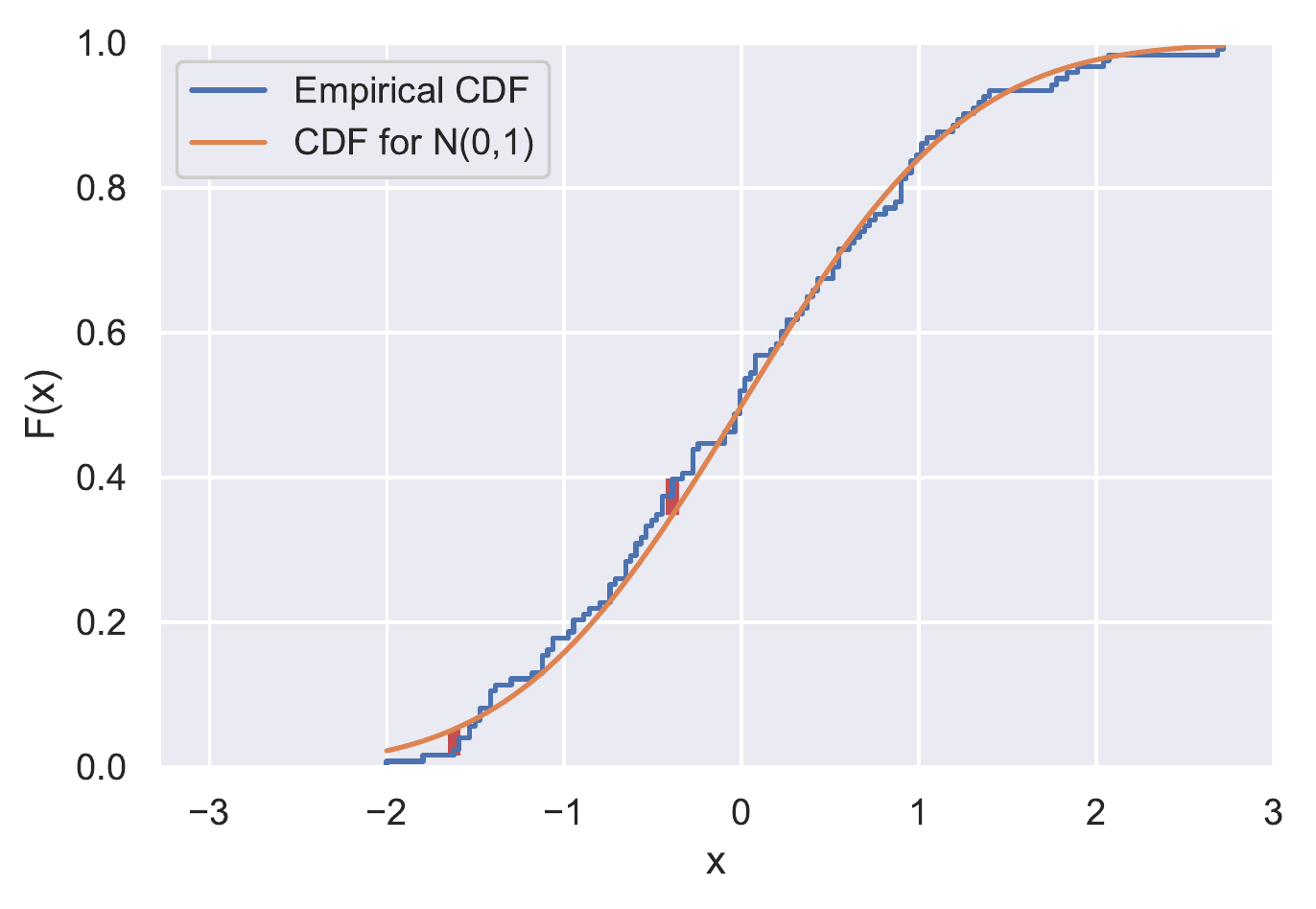} \\
   (a) S\textunderscore6 Tram Station   & (b) S\textunderscore6's CDF  \\[6pt]
 \hspace{0cm}\includegraphics[width=40mm]{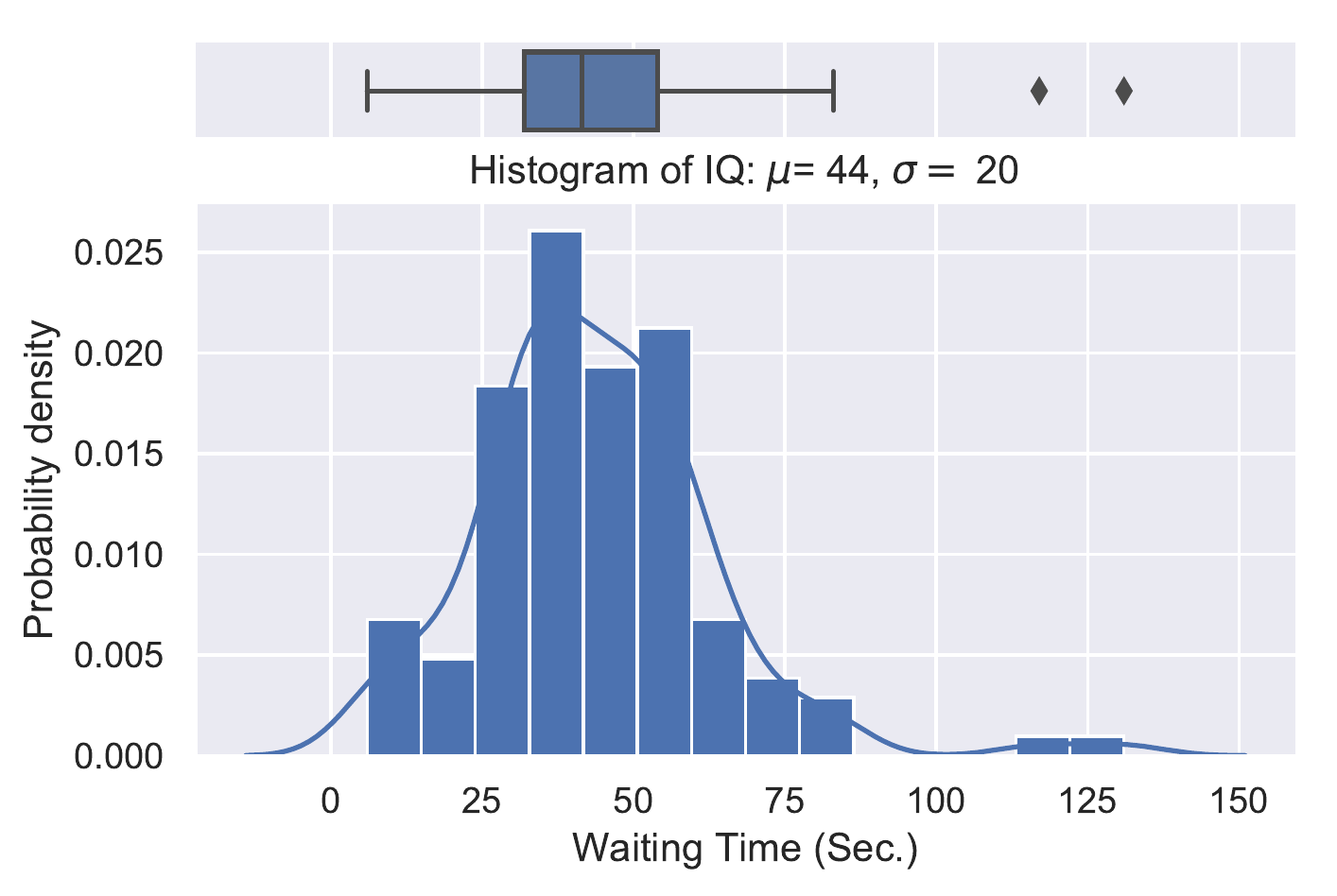} & 
\hspace{0cm}\includegraphics[width=40mm]{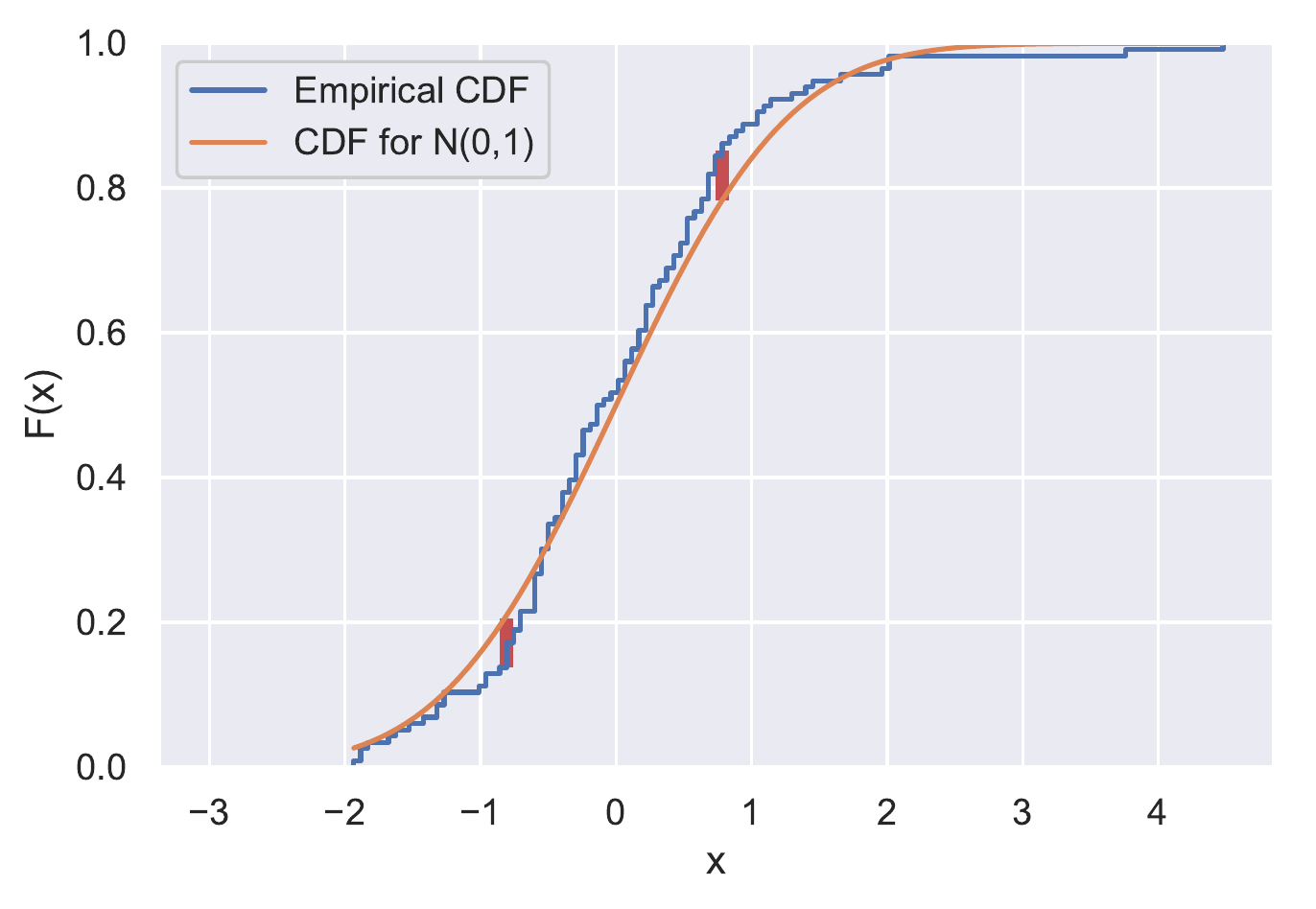} \\
 (c) S\textunderscore7 Tram Station   & (d) S\textunderscore7 CDF   \\[6pt]
\hspace{0cm}\includegraphics[width=40mm]{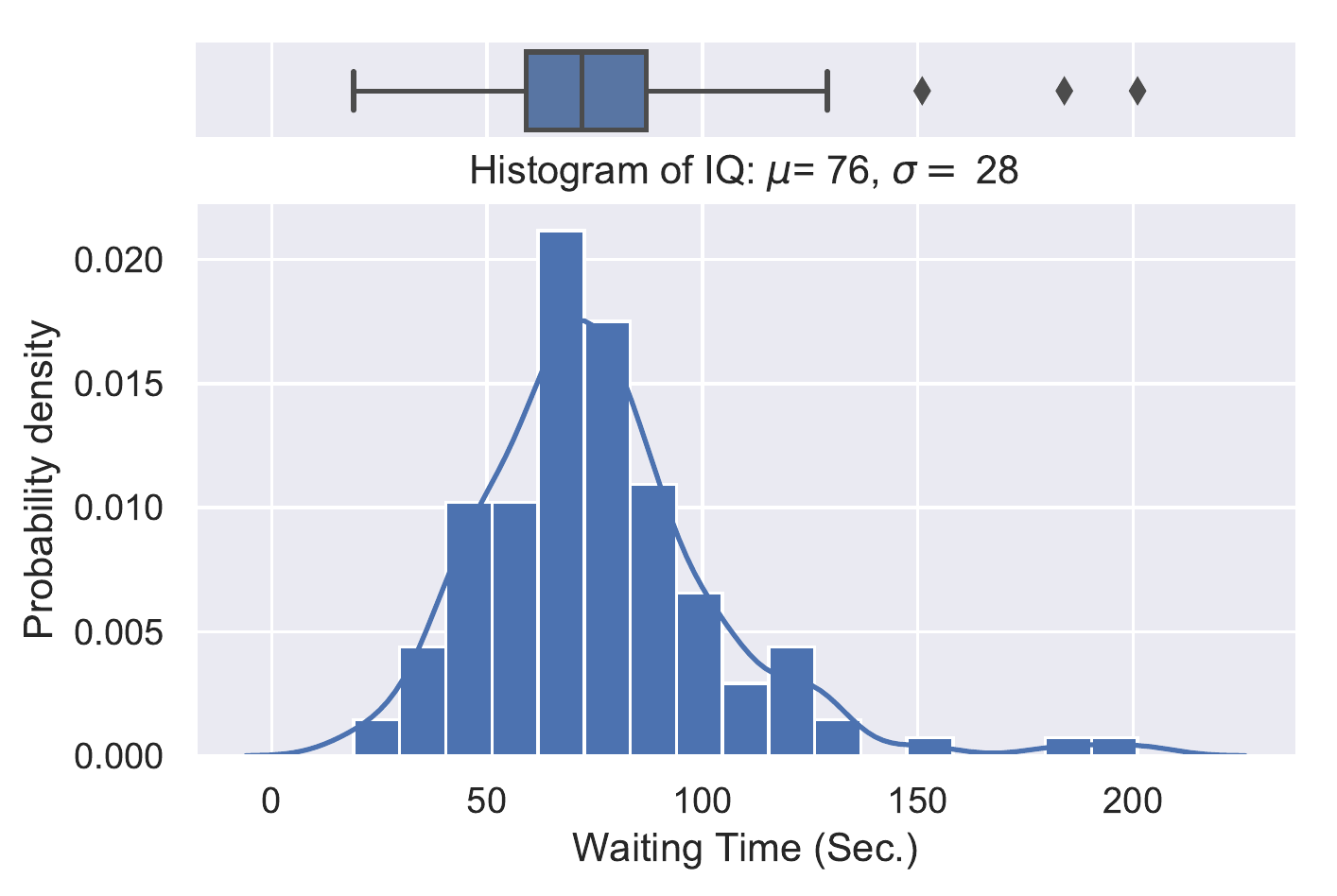} & 
\hspace{0cm}\includegraphics[width=40mm]{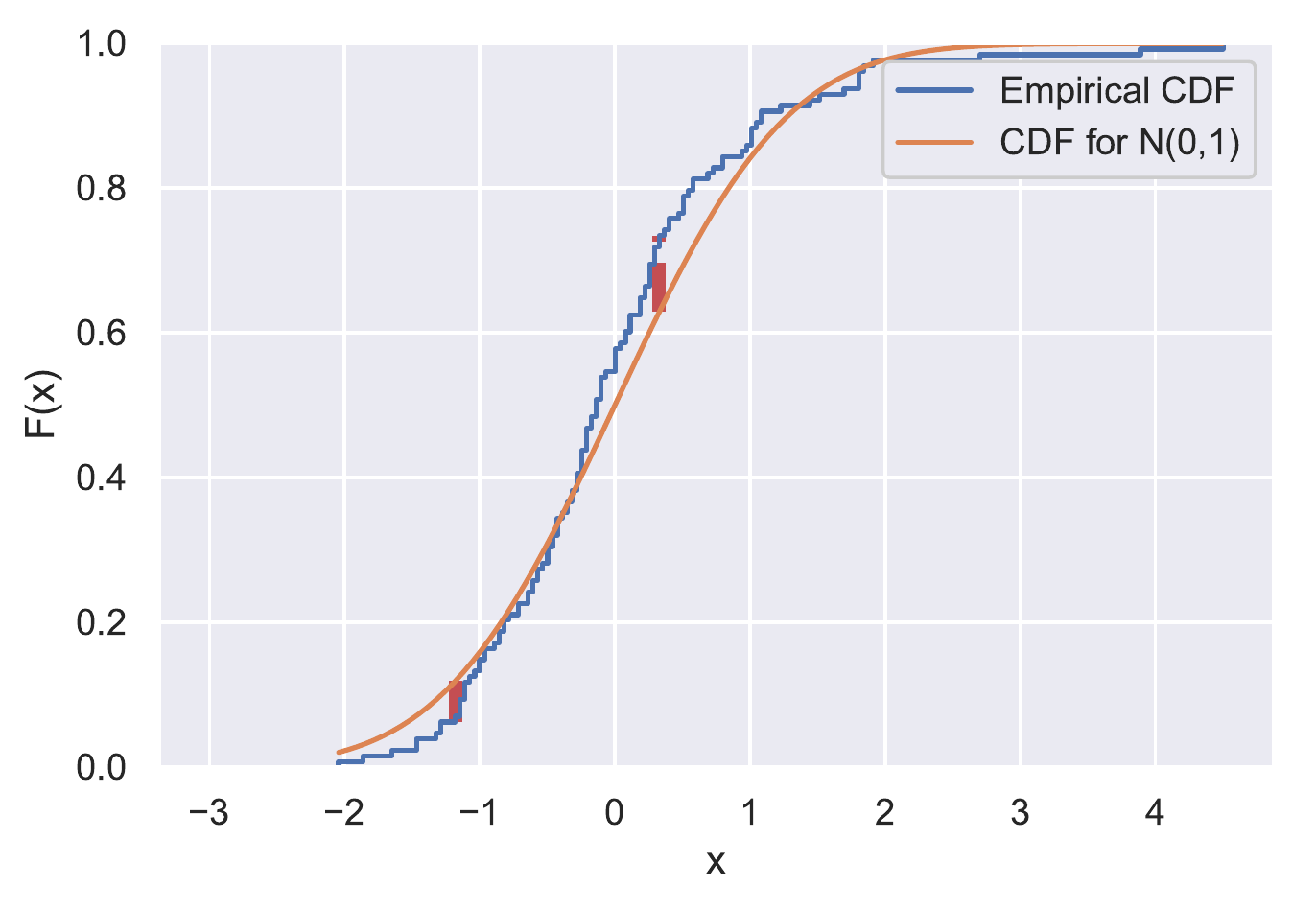} \\
(e) S\textunderscore8 Tram Station   & (f) S\textunderscore8 CDF     \\[6pt]
\end{tabular}
\caption{Waiting time distribution over a long time for several tram stations.}
\label{WaitingTimesInsideTramStations}
\end{figure}

Figure~\ref{WaitingTimesInsideTramStations} shows the  the PDF (propabilty density function) of the waiting time $\evt{w_{s}}$ random variable for three different stations as well as their CDFs. The KsTest\footnote{One-sample Kolmogorov-Smirnov test, which returns a test decision for the null hypothesis that the data in random variable x comes from a standard normal distribution, against the alternative that it does not come from such a distribution}  has been conducted successfully for normality check.
The red marks show the  maximum absolute difference between the calculated and the hypothesized CDFs based on the following equation.
\begin{equation}\label{maximumDifferenceSamir}
D^{*}=\stackunder{max}{x}(| \widehat{f}(x)-G(x)|)
\end{equation}
where $\widehat{f}(x)$ is the empirical CDF and G(x) is the CDF of the hypothesized distribution.
Station waiting times $\evt{w_{s}}$ as well as travel times $\evt{w_{sg}}$ affect a whole trip schedule. If an instance of a segment time duration $w_{sg} > \evt{w_{sg}}$, this implies that an extra delay of an amount of  $w_{sg}-\evt{w_{sg}}$ is expected. To estimate  $\evt{w_{sg}}$, we have collected history data for a long period of time. 
\begin{figure}
\vspace{-0.5cm}
\includegraphics[width=80mm]{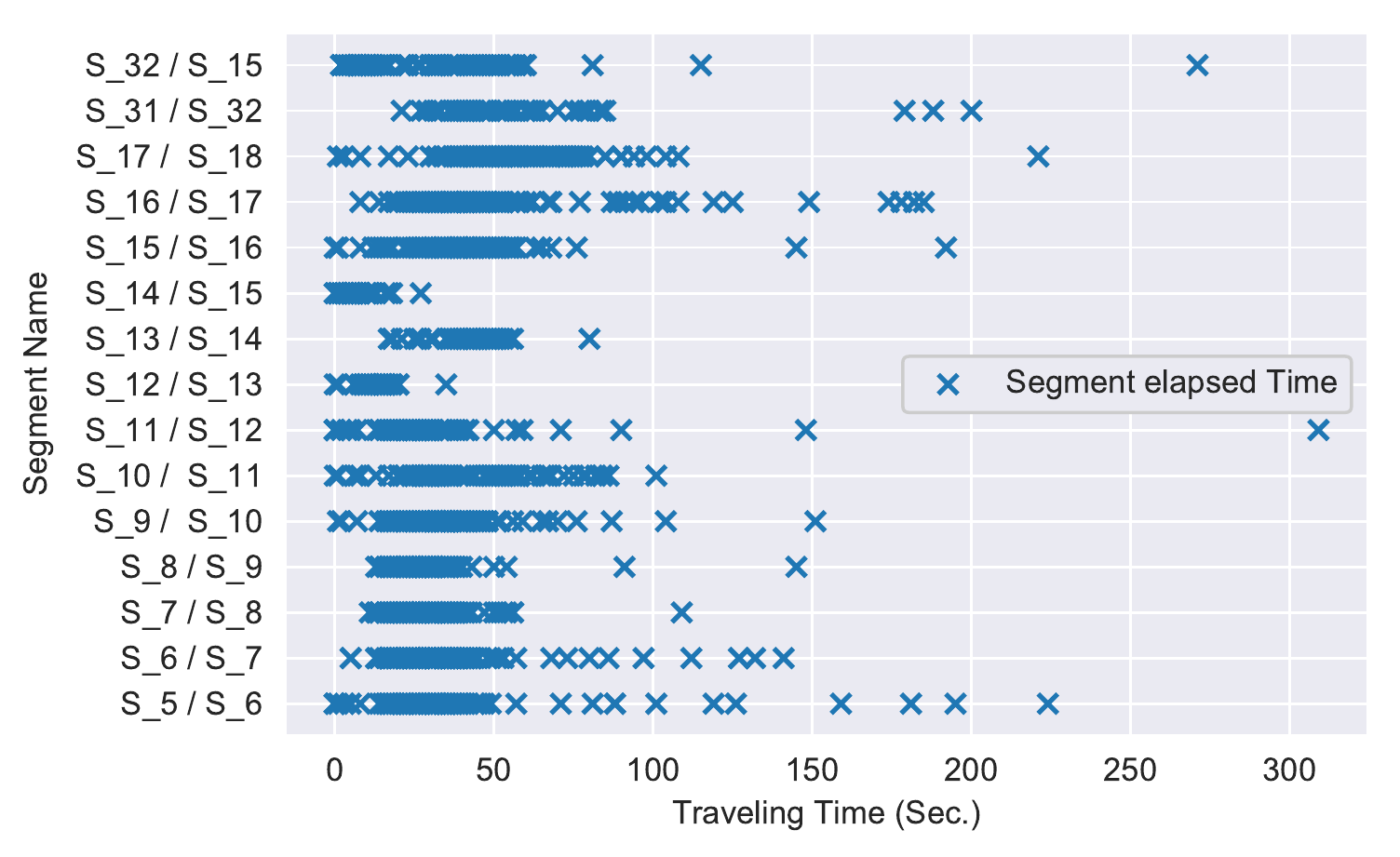}
\caption{ Travel time observations between successive stations ($w_{sg} $) or between stations and traffic lights ($w_{lg}$). }
\label{SegmentTimesAll}
	 \vspace{-0.7cm}
\end{figure}
\begin{figure}
\begin{tabular}{cc}
\hspace{0cm}\includegraphics[width=40mm]{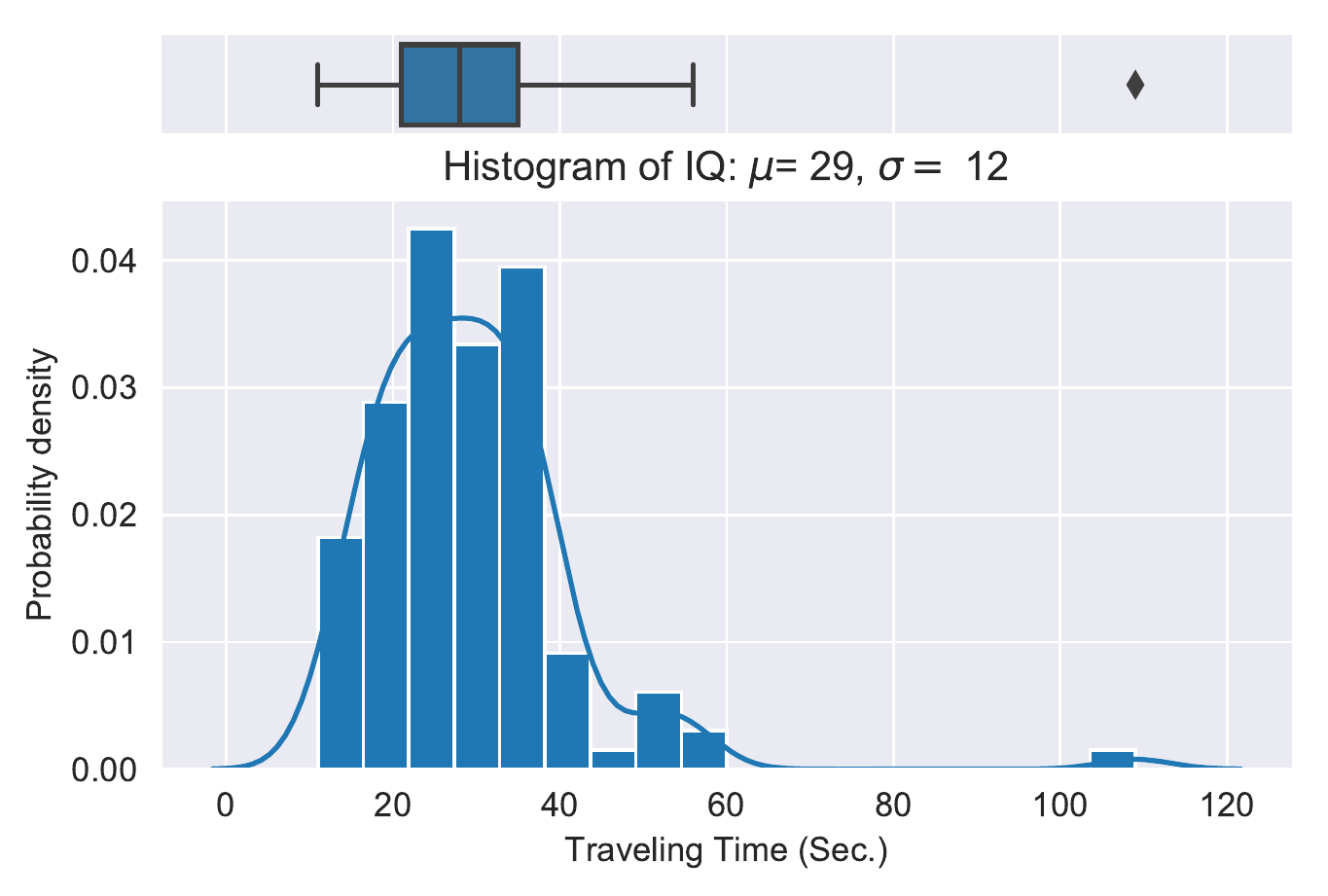} &
\hspace{0cm}\includegraphics[width=40mm]{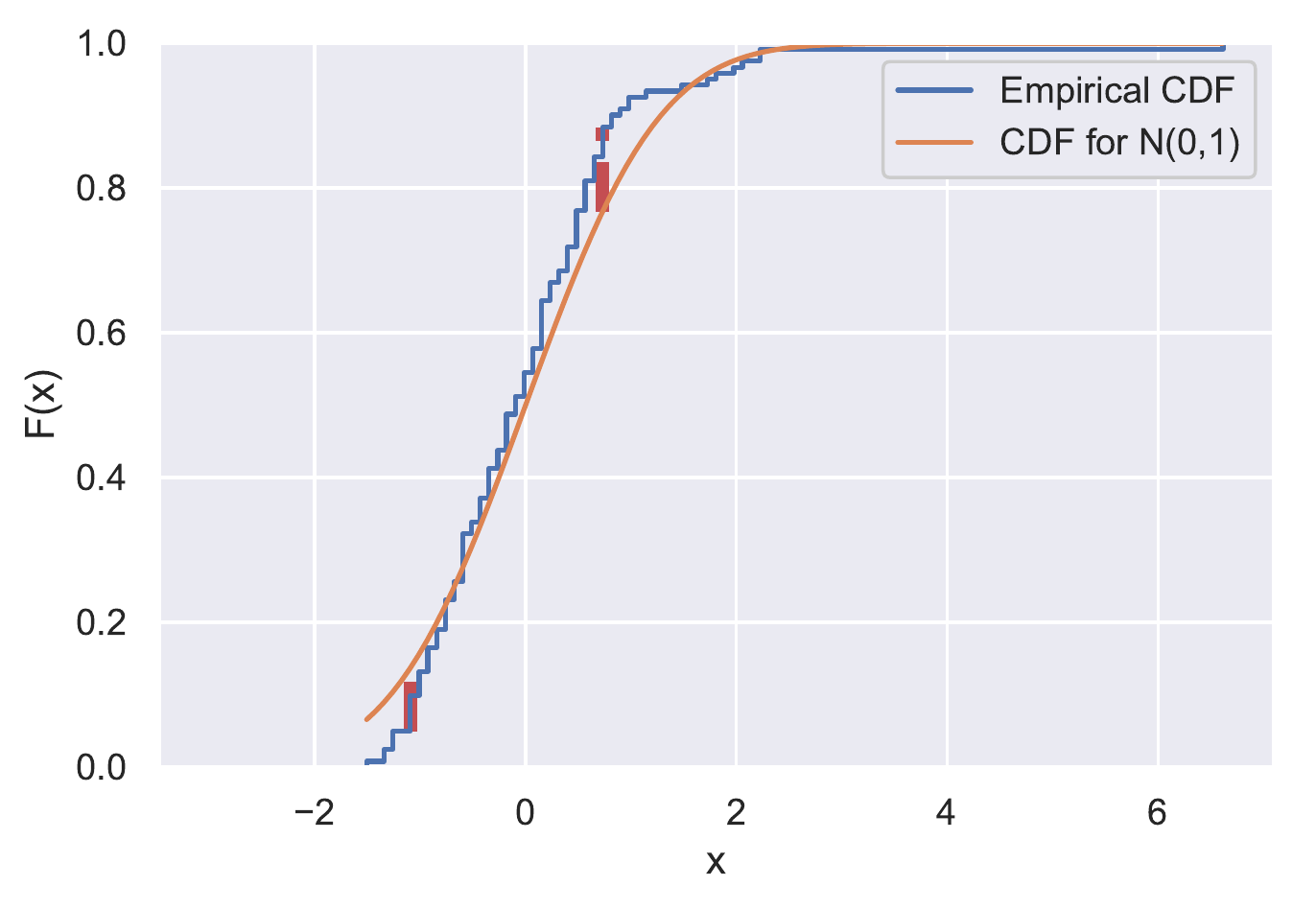} \\
   (a) \small{S\textunderscore7 to S\textunderscore8}  & (b) \small{S\textunderscore7 to S\textunderscore8 CDF}  \\[6pt]
 \hspace{0cm}\includegraphics[width=40mm]{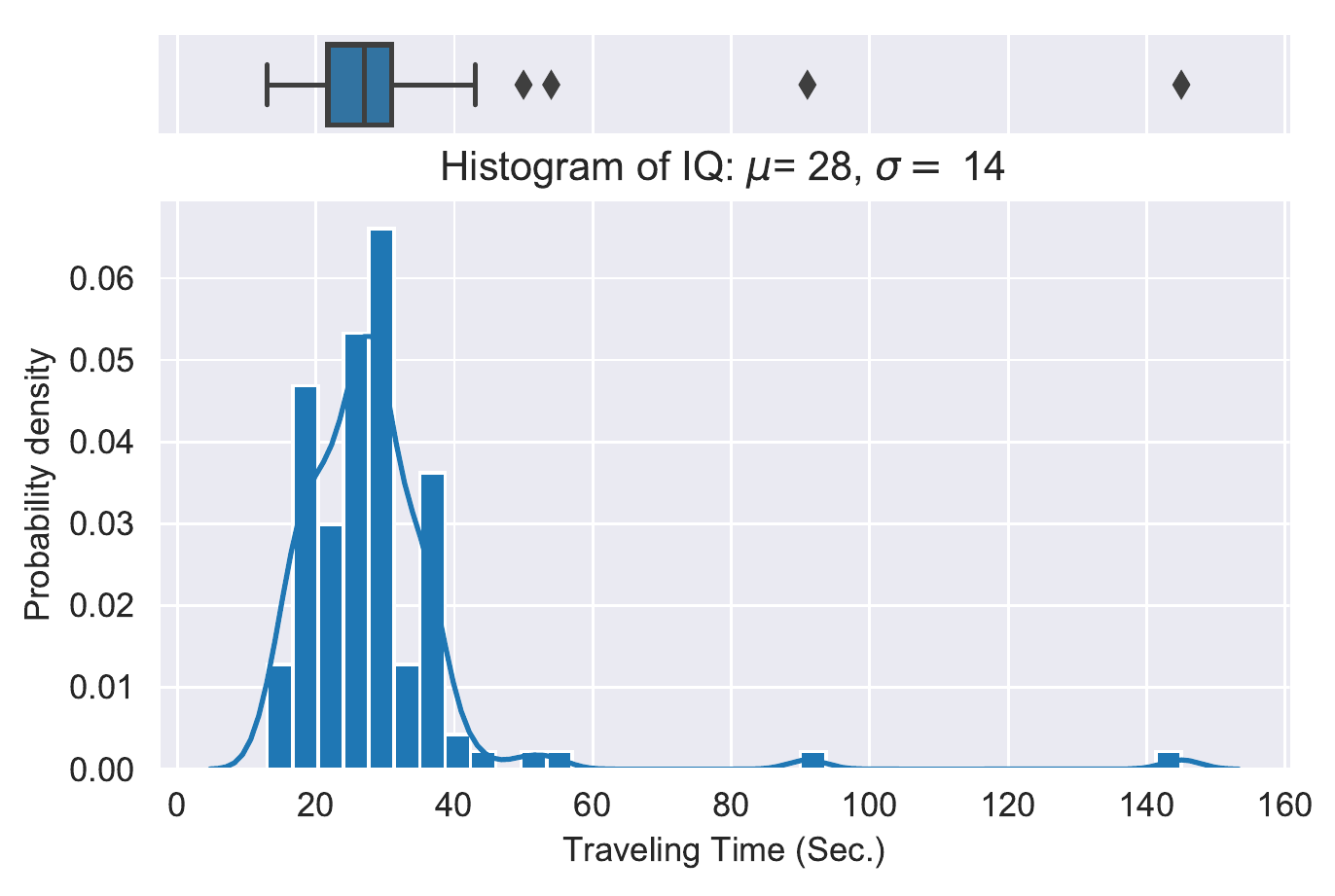} & 
\hspace{0cm}\includegraphics[width=40mm]{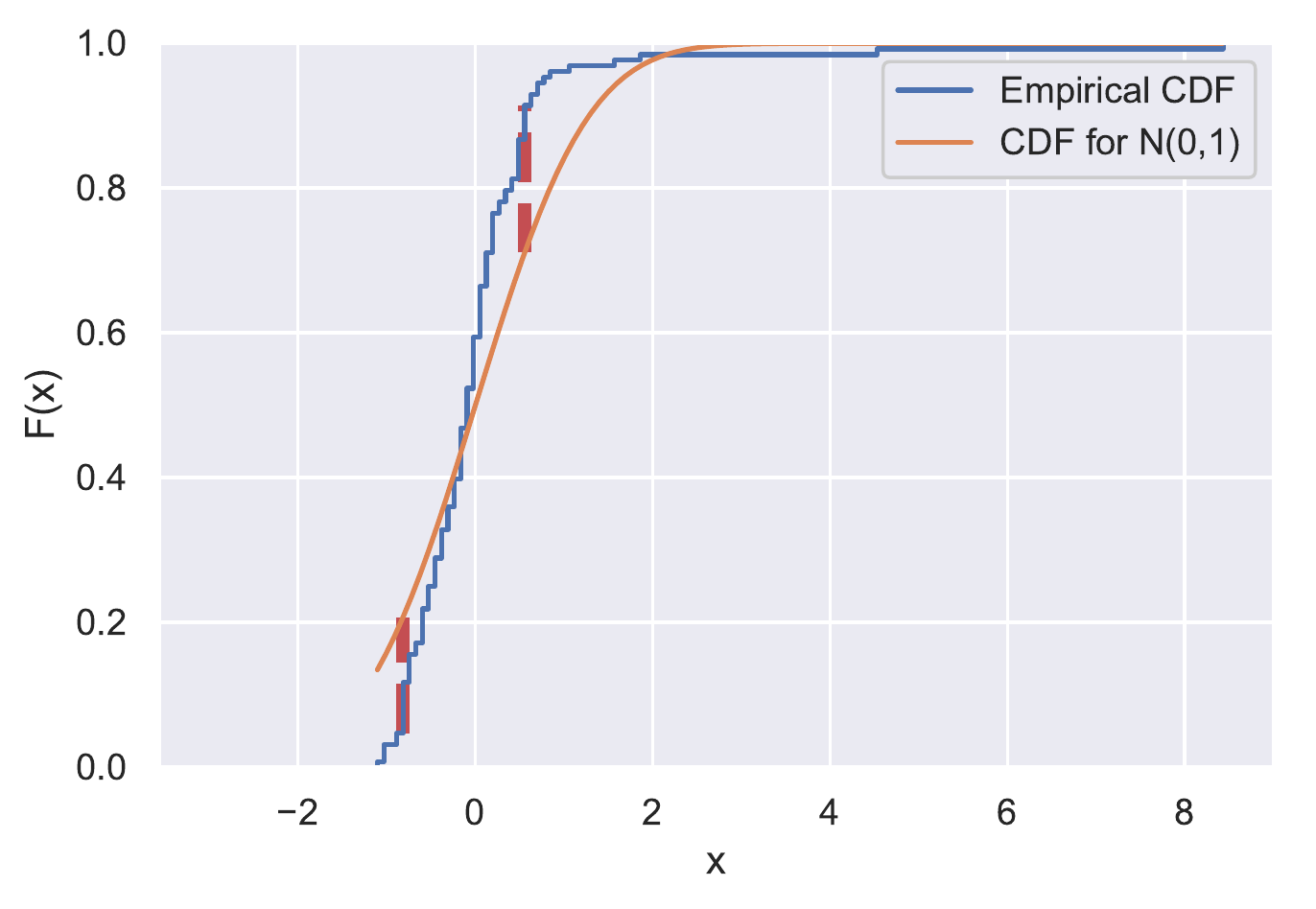} \\
 (c) \small{S\textunderscore8  to S\textunderscore9 }   & (d) \small{S\textunderscore8  to S\textunderscore9 CDF}   \\[6pt]
\hspace{0cm}\includegraphics[width=40mm]{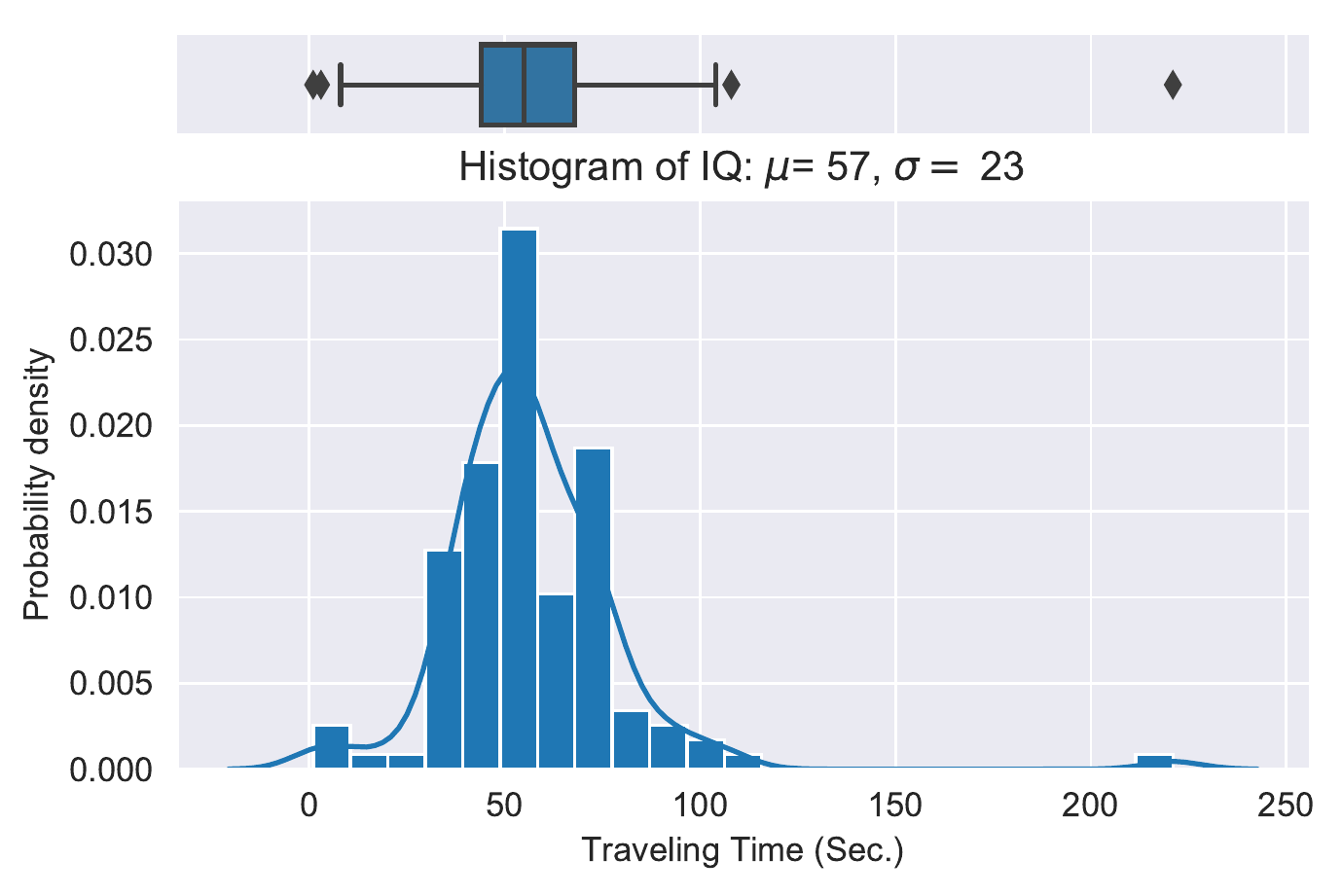} & 
\hspace{0cm}\includegraphics[width=40mm]{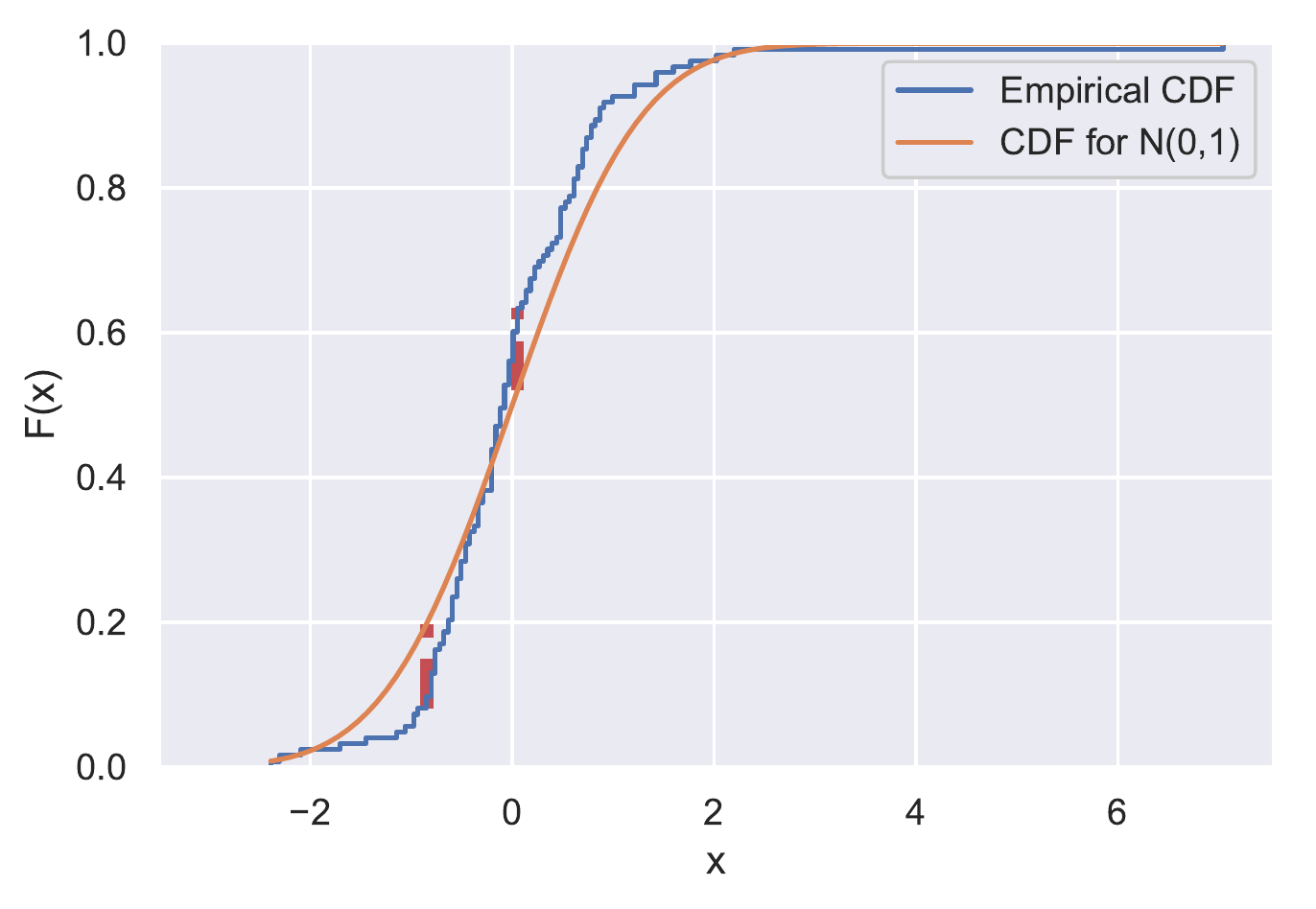} \\
(e) \small{S\textunderscore17 to S\textunderscore18}  & (f) \small{S\textunderscore17 to S\textunderscore18 CDF}    \\[6pt]
\end{tabular}
\caption{Travel time distribution between different successive stations in the given LRT system.}
\label{SegmentTimesInTramforSpecificStations}
	 \vspace{-0.7cm}
\end{figure}
Figure~\ref{SegmentTimesAll} plots samples of the traveling time between consecutive stations without traffic lights ($\evt{w_{sg}}$) interruptions or with traffic lights interruptions ($\evt{w_{lg}}$). Moreover, Figure~\ref{SegmentTimesInTramforSpecificStations} a, c and b, d show the PDF and CDF of ($\evt{w_{sg}}$) between consecutive stations without traffic lights  interruptions. However, Figure~\ref{SegmentTimesInTramforSpecificStations} e, f show the same distribution between two consecutive given stations with traffic light interruption ($\evt{w_{lg}}$). 
Since the random parameters $w_{s}$, $w_{sg}$, $w_{f}$, $w_{bf}$ and $w_{lg}$ have been empirically found to follow normal distributions with means $\mu_{s}$, $\mu_{sg}$, $\mu_{f}$, $\mu_{bf}$, $\mu_{lg}$  and variances  ${\sigma^2}_{s}$, ${\sigma^2}_{sg}$, ${\sigma^2}_{f}$, ${\sigma^2}_{bf}$, ${\sigma^2}_{lg}$,  respectively, a trip time follows a normal distribution of the form: 
\begin{equation}\label{SumOfNormalDist}
N\left(\sum\limits_{i=1}^n c_i \mu_i,\sum\limits_{i=1}^n c^2_i \sigma^2_i\right)
\end{equation}
If a passenger rides a tram from S\textunderscore7  station (source) heading S\textunderscore9 station (destination), the expected trip time would be:
\begin{equation} \label{mueq1}
\begin{split}
\evt{w_{t}}&=\mu_{S\textunderscore7 Station}+\mu_{S\textunderscore7/S\textunderscore8 Segment}+\mu_{S\textunderscore8 Station}\\
&+\mu_{S\textunderscore8/S\textunderscore9 Segment}\\
&=44+29+76+28=177~seconds
\end{split}
\end{equation}
 with standard error of:
\begin{equation} \label{sigmaeq1}
\begin{split}
{\sigma}_{total}&=({\sigma^2}_{S\textunderscore7 Station}+{\sigma^2}_{S\textunderscore7/S\textunderscore8 Segment}\\
&+{\sigma^2}_{S\textunderscore8 Station}+{\sigma^2}_{S\textunderscore8/S\textunderscore9 Segment})^{\frac{1}{2}}\\
&=\sqrt{20^2+12^2+34^2+14^2}=14.5~seconds.
\end{split}
\end{equation}
Note that in Eq. \ref{sigmaeq1} the waiting time at the destination station should not be taken into consideration. The  \Transsense{} estimates the arrival time with an accuracy of 91.81\%.  


	\subsection { History Component.}\label{Sec:HistoryComponent}
	At initalization time, the system is bootstrapped using historical data. Historical data has been collected from many users over a long time duration. This data is useful in estimating and storing the random parameters  $\mu_{s}$, $\mu_{sg}$, $\mu_{f}$, $\mu_{bf}$, $\mu_{lg}$  and variances  ${\sigma^2}_{s}$, ${\sigma^2}_{sg}$, ${\sigma^2}_{f}$, ${\sigma^2}_{bf}$ and ${\sigma^2}_{lg}$ characterizing the obtained distributions for the investigated random parameters.  Because of the increase in population and traffic densities as well as possible service upgrade, there is a need to update the former parameters by a kind of stochastic adaptation (incremental learning)\footnote{ The exact strategy of this type of learning is under investigation.}.


\subsection {Tram ID and Direction Detection Component.}\label{Sec:TramIDandDirectionDetectionComponent}
In the adopted LRT system, there are only two possible directions for a tram: from east to west or from west to east. A tram route direction can be estimated by getting the difference between two sets of GPS readings. As for the tram ID, it can be identified using map matching \cite{semMatch,AccurateRealtimeMapMatching} at splitting areas (see Figure~\ref{fig:AlexandriaTram}).






\section{Related work}

Special devices have been installed in  transporation vehicles to estimate the tranportation system parameters. This includes GPS devices or specific smart phones attached to the tracked transportation vehicles. However, independent GPS devices are relatively expensive and using a small number of smart phones (typically one per vehicle) does not provide enough data to estimate the required variables. 

\cite{Howlongtowait} experimented with a lightweight system for the prediction of bus arrival times that does not need any GPS equipment nor GPS-enabled mobile phones. Instead, it relies on using commodity mobile phones to sense the nearby celltower IDs and to record the beep audio responses, of the IC transit card readers deployed for collecting bus fees. A beep sound is used to confirm that a passenger is in a bus, while a celltower sequence is associated with a bus route. 

 EasyTracker~\cite{EasyTracker} enables bus tracking by analysing GPS traces collected from installed mobile phone in each bus. This requires a lot of time to converge. In addition, they cannot be applied to vehicles that go inside tunnels. This is not the case for \Transsense{} that can identify the get on/get off sequences using the passenger behaviour. 

The UrbanEye system \cite{UrbanEye} describes some peculiarities of bus transit systems especially in developing countries. These include chaotic stoppage patterns and unpredictable speed variations. However, analyzing lightrail systems can lead to better results as their motion and stopping patterns are more predictable.

Estimating the location of a cellular phone or transportation vehicle can be based on different sensors including GPS, sensor-based landmarks \cite{aly2013dejavu, AccurateRealtimeMapMatching}, or cellular signals \cite{CellSense, WiDeep, Monosense, Monosense2}. In this paper, we leveraged the GPS system in smart phones. However, other localization systems with comparable accuracy to GPS and lower energy-consumption, e.g., can also be used.

\section{Conclusion}\label{Sec:conclusion}

This paper proposes \Transsense{} that estimates the expected waiting time of a passenger for getting on/ off to/ from a LRT.  \Transsense{} components are explained in detail.
The basic idea is that a phone speed can be correlated with the GPS data to extract the semantics,  hidden in the filtered samples, such as the average tram waiting time at stations, traffic lights and travel time between stations. This makes it possible to construct a  real time tram schedule for the investigated system. Over 800 hours of daily passengers' traces have been collected using different tram lines at different time periods. \Transsense{} achieved an average recall and precision of 95.35\% and  90.1\%, respectively, in  discriminating  between  stations  and traffic lights. Moreover, Trans-Sense is able to calculate the  stations’  dimensions with  an  accuracy  of 95.714\% and can incorporate more stations based only on the information provided from GPS. The system estimates the right time of arrival with an accuracy of 91.81\%.

 The \Transsense{} working scenario, decribed in this paper, can be easily applied to  trains/ metros whose stopping patterns are predictable as well as similar lightrail transportation systems with larger number of lines/ routes. Generalization to vehicles with chaotic  stoppage  patterns as well as the exact strategy of the stochastic learning for the adaptation of the learned distributions are future research topics.

\bibliographystyle{IEEEtran}
\bibliography{Traffic15}

\end{document}